\documentclass[11pt]{article}
\usepackage[utf8]{inputenc}
\usepackage[T1]{fontenc}
\usepackage{tensor}
\usepackage{bbm}
\usepackage{amssymb}
\usepackage{authblk}
\usepackage{xcolor}
\usepackage{graphicx}
\usepackage{jheppub}
\usepackage{bm}

\usepackage[multiple]{footmisc}
\usepackage{subcaption}
\usepackage{multicol}
\usepackage{multirow}
\usepackage{rotating}
\usepackage{array}
\usepackage{longtable}
\usepackage{booktabs}
\usepackage[mathscr]{euscript}
\usepackage{tikz}
\usepackage{tikz-cd}
\usepackage{braket}
\usepackage{dsfont}

\tikzset{
	partial ellipse/.style args={#1:#2:#3}{
		insert path={+ (#1:#3) arc (#1:#2:#3)}
	}
}

\tikzset{snake it/.style={decorate, decoration=snake}}

\usetikzlibrary{calc,decorations.markings}
\usetikzlibrary{decorations.pathmorphing}
\usetikzlibrary{decorations.pathreplacing}
\usetikzlibrary{shapes,backgrounds}
\usetikzlibrary{fadings}
\usetikzlibrary{cd}
\usetikzlibrary{hobby}
\usetikzlibrary{decorations.pathreplacing}

\tikzfading
[
	name=fade out,
	inner color=transparent!0,
	outer color=transparent!100
]

\newcommand{\be}{ \begin{equation}}
\newcommand{\ee}{\end{equation}}

\newcommand{\bket}[1]{\| #1 \rangle\hspace{-0.075cm}\rangle}

\makeatletter
\newcommand*\bigcdot{\mathpalette\bigcdot@{.65}}
\newcommand*\bigcdot@[2]{\mathbin{\vcenter{\hbox{\scalebox{#2}{$\m@th#1\bullet$}}}}}
\makeatother

\title{\boldmath Classical geometry from the tensionless string}

\author{Bob Knighton}

\affiliation{Institut f\"ur Theoretische Physik, ETH Z\"urich \\
\hspace*{0.3cm} Wolfgang-Pauli-Strasse 27, 8093 Z\"urich, Switzerland}

\emailAdd{robejr@ethz.ch}

\abstract{Tensionless string theory on $\text{AdS}_3\times\text{S}^3\times\mathcal{M}$ is explored in the limit that the strings wind the asymptotic boundary a large number of times. Although the worldsheet is usually thought to be localised to the $\text{AdS}$ boundary, we argue that the string can actually probe the bulk geometry in this limit. In particular, we show that correlation functions can be expressed in terms of a minimal-area worldsheet propagating in $\text{AdS}_3$. We then relate the classical motion of the string to the twistor-like free field description of the tensionless worldsheet theory. Finally, we consider a particular dimensional reduction of $\text{AdS}_3$ to $\text{AdS}_2$, and show that the effective action of the worldsheet formally resembles the one-dimensional Schwarzian theory of JT gravity with conical defects.}

\begin{document}

\maketitle

\section{Introduction}

The AdS/CFT correspondence relates a UV-complete theory of quantum gravity on Anti-de Sitter spacetime in $d$ dimensions to a superconformal field theory in $d-1$ dimensions. In the limit in which the string tension is taken to be large, the AdS gravity theory is governed by the semiclassical limit of supergravity, and the dual theory is described by a strongly-coupled field theory in one dimension lower. This duality then allows one to calculate certain quantities in strongly-coupled field theories using semiclassical general relativity. On the other end of the spectrum of holographic dualities, one can consider a weakly coupled conformal field theory, which should be dual to a theory for which the characteristic radius of the spacetime is on the string scale. Since, in this limit, one cannot trust the semiclassical treatment of the AdS theory in terms of semiclassical geometry, one must rely on the worldsheet approach to string theory in order to perform any trustworthy calculations.

In recent years, this limit of the AdS/CFT correspondence, dubbed the \textit{tensionless} limit, has been shown to be surprisingly rich and possesses the rare quality that both sides of the duality are analytically tractable. In particular, the tensionless ($k=1$) limit of string theory on $\text{AdS}_3\times\text{S}^3\times\mathcal{M}$ with pure NS-NS flux, where $\mathcal{M}$ is either $\mathbb{T}^4$ or $\text{K}3$, has a precisely-known perturbative CFT dual, given by the symmetric orbifold theory $\text{Sym}^K(\mathcal{M})$ of the 2D sigma model on the internal manifold $\mathcal{M}$. The duality between the symmetric orbifold theory and the tensionless string has passed a large number of consistency checks, including a matching of the full (single-particle) partition function and (connected) correlation functions on both sides \cite{Gaberdiel_2018,Eberhardt:2018,Eberhardt:2019,Dei:2020,Eberhardt:2020,Knighton:2020kuh,Bertle:2020sgd,Gaberdiel:2021njm,Eberhardt:2020bgq}.

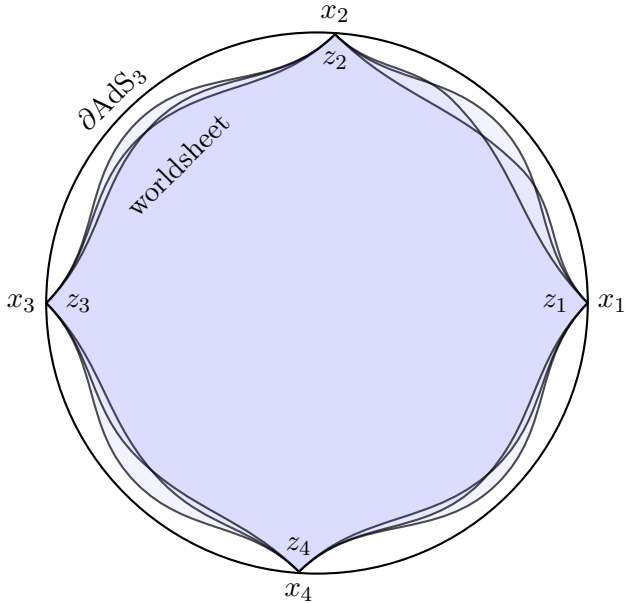
\begin{figure}
\centering
\begin{tikzpicture}[scale = 1.2]
\draw[thick] (0,0) circle (3);
\path[draw, closed=true, thick, black, opacity = 0.7] (3,0) to[out = 135, in = -45] (2.25,1.5) to[out = 135, in = -45] (0.2,2.98) to[out = -135, in = 45] (-2,1.8) to[out = -135, in = 45] (-3,0) to[out = -45, in = 135] (-1.7,-2) to[out = -45, in = 135]  (-0.2,-2.98) to[out = 45, in = -135] (2,-1.8) to[out = 45, in = -135] (3,0);
\path[fill, closed=true, thick, blue, opacity = 0.05] (3,0) to[out = 135, in = -45] (2.25,1.5) to[out = 135, in = -45] (0.2,2.98) to[out = -135, in = 45] (-2,1.8) to[out = -135, in = 45] (-3,0) to[out = -45, in = 135] (-1.7,-2) to[out = -45, in = 135]  (-0.2,-2.98) to[out = 45, in = -135] (2,-1.8) to[out = 45, in = -135] (3,0);
\path[draw, closed=true, thick, black, opacity = 0.7] (3,0) to[out = 135, in = -45] (2,2) to[out = 135, in = -45] (0.2,2.98) to[out = -135, in = 45] (-2,2) to[out = -135, in = 45] (-3,0) to[out = -45, in = 135] (-2,-2) to[out = -45, in = 135]  (-0.2,-2.98) to[out = 45, in = -135] (2,-2) to[out = 45, in = -135] (3,0);
\path[fill, closed=true, thick, blue, opacity = 0.05] (3,0) to[out = 135, in = -45] (2,2) to[out = 135, in = -45] (0.2,2.98) to[out = -135, in = 45] (-2,2) to[out = -135, in = 45] (-3,0) to[out = -45, in = 135] (-2,-2) to[out = -45, in = 135]  (-0.2,-2.98) to[out = 45, in = -135] (2,-2) to[out = 45, in = -135] (3,0);
\path[draw, closed=true, thick, black, opacity = 0.7] (3,0) to[out = 135, in = -45] (1.5,2.25) to[out = 135, in = -45] (0.2,2.98) to[out = -135, in = 45] (-1.8,2) to[out = -135, in = 45] (-3,0) to[out = -45, in = 135] (-2,-1.7) to[out = -45, in = 135]  (-0.2,-2.98) to[out = 45, in = -135] (1.8,-2) to[out = 45, in = -135] (3,0);
\path[fill, closed=true, thick, blue, opacity = 0.05] (3,0) to[out = 135, in = -45] (1.5,2.25) to[out = 135, in = -45] (0.2,2.98) to[out = -135, in = 45] (-1.8,2) to[out = -135, in = 45] (-3,0) to[out = -45, in = 135] (-2,-1.7) to[out = -45, in = 135]  (-0.2,-2.98) to[out = 45, in = -135] (1.8,-2) to[out = 45, in = -135] (3,0);
\node[right] at (3,0) {$x_1$};
\node[left] at (2.9,0) {$z_1$};
\node[above] at (0.2,3) {$x_2$};
\node[below] at (0.2,2.9) {$z_2$};
\node[left] at (-3,0) {$x_3$};
\node[right] at (-2.9,0) {$z_3$};
\node[below] at (-0.2,-3) {$x_4$};
\node[above] at (-0.2,-2.9) {$z_4$};
\node[below, rotate = 45] at (-1.7,1.7) {worldsheet};
\node[above, rotate = 45] at (-2.1,2.1) {$\partial\text{AdS}_3$};
\end{tikzpicture}
\caption{A worldsheet representing the correlation functions of operators at positions $\{x_i\}$ in the CFT dual. The strings are localised to the $\text{AdS}_3$ boundary and the worldsheet holomorphically covers the sphere.}
\label{fig:glued-worldsheet}
\end{figure}

The tractability of this instance of the AdS/CFT duality comes from the rather remarkable fact that the worldsheet path integral localises to worldsheet configurations for which the string is `glued' to the $\text{AdS}_3$ boundary \cite{Eberhardt:2019}, see Figure \ref{fig:glued-worldsheet}.\footnote{This is related to the fact that in the tensionless limit of string theory on $\text{AdS}_3\times\text{S}^3\times\mathcal{M}$, the only representations of $\mathfrak{psu}(1,1|2)_1$ which appear in the worldsheet partition function correspond to the `long strings' of \cite{Maldacena-Ooguri-1}.} For these configurations, one can consider the worldsheet as covering the $\text{AdS}_3$ boundary. Indeed, one can show that the only worldsheet configurations which contribute in the string theory path integral arise from holomorphic covering maps $\Gamma:\Sigma\to\partial(\text{AdS}_3)$ which are branched at the appropriate insertion points \cite{Eberhardt:2019,Dei:2020,Knighton:2020kuh,Eberhardt:2021jvj}. These covering spaces are precisely those which are considered in the construction of correlation functions of twist fields in the symmetric orbifold theory \cite{Lunin_2001,Dei:2019}, and thus the localisation of the string theory path integral to these covering spaces provides the fundamental mechanism for this particular AdS/CFT correspondence.

Although the duality between the tensionless string theory on $\text{AdS}_3\times\text{S}^3\times\mathbb{T}^4$ and the symmetric orbifold theory is computationally well understood, it is somewhat lacking in geometric interpretation. In particular, a description of the bulk degrees of freedom is almost entirely absent. The reason for this is two-fold. First, as the name suggests, the tensionless limit of string theory is constructed in the limit where the fundamental strings have a size comparable to that of the radius of curvature of the background geometry, and thus a geometric description of the bulk in terms of local fields (which typically emerge from the point-particle limit of string theory) cannot be expected to exist. Second, because the string path integral only includes contributions from strings which lie near the boundary, the exact details of the bulk geometry (if one can be thought to exist) are completely invisible to the worldsheet. Indeed, it has been shown that correlation functions and partition functions of the tensionless string are background-independent, in the sense that two bulk geometries give rise to the same partition functions, so long as they have the same asymptotic boundaries \cite{Eberhardt:2021jvj} (at least for a large class of bulk geometries).

It would therefore seem hopeless to expect a reasonable bulk interpretation of the tensionless string theory, and that in order to pose meaningful questions about the bulk geometry one would need to deform away from the tensionless limit. However, as we will show, the bulk geometry of $\text{AdS}_3$ can in fact be seen from the point of view of the worldsheet in the limit that the number of times the worldsheet covers the boundary (i.e. the degree of the holomorphic map $\Gamma:\Sigma\to\partial(\text{AdS}_3)$) becomes very large, which we will refer to as the \textit{large-twist limit}. This limit has been previously considered in the literature from two separate perspectives:

\subsubsection*{Stringy geometry}

If one considers the `grand canonical' symmetric orbifold theory
\begin{equation}
\text{Sym}(\mathbb{T}^4)=\bigoplus_{K=0}^{\infty}\text{Sym}^K(\mathbb{T}^4)\,,
\end{equation}
then one can define the grand canonical partition function on a Riemann surface $\Sigma$ to be
\begin{equation}
\mathfrak{Z}(p,\Sigma)=\sum_{K=0}^{\infty}p^KZ_{\text{Sym}^K(\mathbb{T}^4)}(\Sigma)\,.
\end{equation}
It was argued in \cite{Eberhardt:2021jvj} that $\mathfrak{Z}(p,\Sigma)$ has poles in the chemical potential $p$, and behaves near these poles like
\begin{equation}\label{eq:sum-over-bulks}
\mathfrak{Z}(p,\Sigma)\sim\sum_{\text{bulk geometries }\mathcal{M}_3}\frac{Z(\mathcal{M}_3)}{\sigma+\frac{3i}{\pi}I_{\text{bulk}}(\mathcal{M}_3)}\,,
\end{equation}
where the sum is over a set of bulk geometries $\mathcal{M}_3$ which have boundary $\Sigma$, $I_{\text{bulk}}(\mathcal{M}_3)$ is the gravitational action of $\mathcal{M}_3$, $Z(\mathcal{M}_3)$ is some suitable one-loop determinant, and $p=e^{2\pi i\sigma}$.

The interpretation of this schematic expression is as follows: a pole in the grand canonical partition function occurs when the Taylor series in $p$ diverges. At these values of $p$, the primary contributions to $\mathfrak{Z}(p,\Sigma)$ come from arbitrarily large values of $K$. From the point of view of the worldsheet, $K$ counts the number of times the worldsheet holomorphically covers the boundary $\Sigma$.\footnote{This is actually only true for \textit{connected} worldsheets. For disconnected worldsheets, $K$ counts the sum of the number of times each worldsheet covers the boundary.} Thus, equation \eqref{eq:sum-over-bulks} tells us that bulk geometries $\mathcal{M}_3$ emerge in the limit that the number $K$ of times the worldsheet(s) cover the boundary is arbitrarily large. In the language of \cite{Eberhardt:2021jvj}, the large number of string worldsheets `condense' into classical three-dimensional geometry. Upon condensation, a string which winds a contractible cycle in the bulk $M$ times generates a conical singularity with deficit angle $\theta=2\pi(1-M^{-1})$.

\subsubsection*{The Strebel programme}

From the point of view of the dual symmetric orbifold CFT $\text{Sym}^{K}(\mathcal{M})$, the large-twist limit enters in considering sphere correlation functions of the form
\begin{equation}
\Braket{\prod_{i=1}^{n}\mathcal{O}_i^{(w_i)}(x_i)}\,,
\end{equation}
where $\mathcal{O}^{(w_i)}_i$ is some state in the $w_i$-cycle twisted sector of the symmetric orbifold. The analysis of Lunin and Mathur \cite{Lunin_2001} showed that the connected part of this correlation function can be expressed as a sum over holomorphic maps $\Gamma:\Sigma\to\mathbb{CP}^1$ satisfying ramification conditions at critical points $z_i$ on $\Sigma$ (see equation \eqref{eq:covering-map} below). The precise form is given by
\begin{equation}\label{eq:Lunin-Mathur}
\Braket{\prod_{i=1}^{n}\mathcal{O}^{(w_i)}_i(x_i)}_{\text{c}}=\sum_{\Gamma:\Sigma\to\mathbb{CP}^1}K^{\chi(\Sigma)}e^{-S_{\text{Liouville}}[\Phi_{\Gamma}]}\Braket{\prod_{i=1}^{n}\mathcal{O}_i(z_i)}_{\Sigma}\,,
\end{equation}
i.e. as a sum of seed-theory correlators on covering surfaces $\Sigma$, weighted by a usual $1/K$ factor in the genus expansion times a conformal factor given by the Liouville action of the field
\begin{equation}
\Phi_{\Gamma}=\log|\partial\Gamma|^2\,.
\end{equation}
The degree of $\Gamma$, as a map, is determined via the Riemann-Hurwitz relation by the twists $w_i$ via
\begin{equation}
\text{deg}(\Gamma)=N=1-g+\sum_{i=1}^{n}\frac{w_i-1}{2}\,.
\end{equation}
In \cite{Gaberdiel:2020ycd}, it was noticed that, in the limit that this degree is taken to be infinite, the number of allowable maps $\Gamma$ becomes infinite and the finite sum in \eqref{eq:Lunin-Mathur} condenses into an integration over the moduli space $\mathcal{M}_{g,n}$ of curves of genus $g$. Concretely, the result of the large-twist analysis is that the correlators in the symmetric orbifold take the suggestively stringy form
\begin{equation}\label{eq:strebel-gauge-NG}
\Braket{\prod_{i=1}^{n}\mathcal{O}^{(w_i)}_i(x_i)}_{\text{c}}=\sum_{g=0}^{\infty}g_s^{2g-2}\int_{\mathcal{M}_{g,n}}\mathrm{d}\mu\,e^{-S_{\text{NG}}}\Braket{\prod_{i=1}^{n}\mathcal{O}_i(z_i)}_{\Sigma}\,,
\end{equation}
where we have set $g_s=1/K$. The effective action can be written in terms of a special meromorphic quadratic differential $\varphi$ on $\Sigma$ and takes the form
\begin{equation}\label{eq:intro-ng}
S_{\text{NG}}=\frac{N^2}{4\pi}\int_{\Sigma}|\varphi|\,,
\end{equation}
which is the area of a string with metric $g_{z\bar{z}}=|\varphi|/4$. The constant of proportionality $N^2$ introduces an `effective' tension
\begin{equation}
T_{\text{eff}}=N^2
\end{equation}
to the string. In terms of the covering map $\Gamma$, the quadratic differential $\varphi$ is given by
\begin{equation}
\varphi=-\frac{2}{N^2}S[\Gamma]\,,
\end{equation}
where $S[\Gamma]$ denotes the Schwarzian derivative of the map $\Gamma$. The quadratic differential $\varphi$ is related, in the large-twist limit, to so-called Strebel differentials, which were argued in \cite{Gopakumar:2003ns,Gopakumar:2004qb,Gopakumar:2005fx} to be instrumental in the mechanism whereby free field theories arrange themselves into AdS string theories. In the case of $\text{AdS}_3/\text{CFT}_2$, it was suggested in \cite{Gaberdiel:2020ycd} that it should be possible to think of the Strebel gauge metric $g_{z\bar{z}}=N^2|\varphi|/4$ as the pullback of the $\text{AdS}_3$ metric onto the worldsheet. If this were the case, then \eqref{eq:strebel-gauge-NG} would be thought of as the semiclassical path integral of a string propagating in $\text{AdS}_3$.

\begin{figure}[!ht]
\centering
\begin{tikzpicture}
\begin{scope}[xshift = -4.5cm,yshift = 2cm]
\draw[thick] (-3,-1) -- (3,-1) -- (3,1) -- (-3,1) -- (-3,-1);
\node at (0,0) {Symmetric product CFT};
\end{scope}
\begin{scope}[xshift = 4.5cm,yshift = 2cm]
\draw[thick] (-3,-1) -- (3,-1) -- (3,1) -- (-3,1) -- (-3,-1);
\node at (0,0) {Tensionless strings on $\text{AdS}_3$};
\end{scope}
\begin{scope}[xshift = -4.5cm, yshift = -2cm]
\draw[thick] (-3,-1) -- (3,-1) -- (3,1) -- (-3,1) -- (-3,-1);
\node at (0,0) {Strebel-gauge strings};
\end{scope}
\begin{scope}[xshift = 4.5cm, yshift = -2cm]
\draw[thick] (-3,-1) -- (3,-1) -- (3,1) -- (-3,1) -- (-3,-1);
\node at (0,0) {`Stringy' $\text{AdS}_3$ geometry};
\end{scope}
\draw[thick, latex-latex] (-1.3,2) -- (1.3,2);
\node[above] at (0,2) {AdS/CFT};
\draw[thick, -latex] (-4.5,0.8) -- (-4.5,-0.8);
\node[left] at (-4.5,0) {Large-twist};
\draw[thick, -latex] (4.5,0.8) -- (4.5,-0.8);
\node[right] at (4.5,0) {Large-twist};
\draw[thick, latex-latex] (-1.3,-2) -- (1.3,-2);
\node[above] at (0,-2) {???};
\end{tikzpicture}
\caption{Two separate approaches to studying the `large-twist' limit of the tensionless AdS/CFT correspondence.}
\label{fig:approaches}
\end{figure}
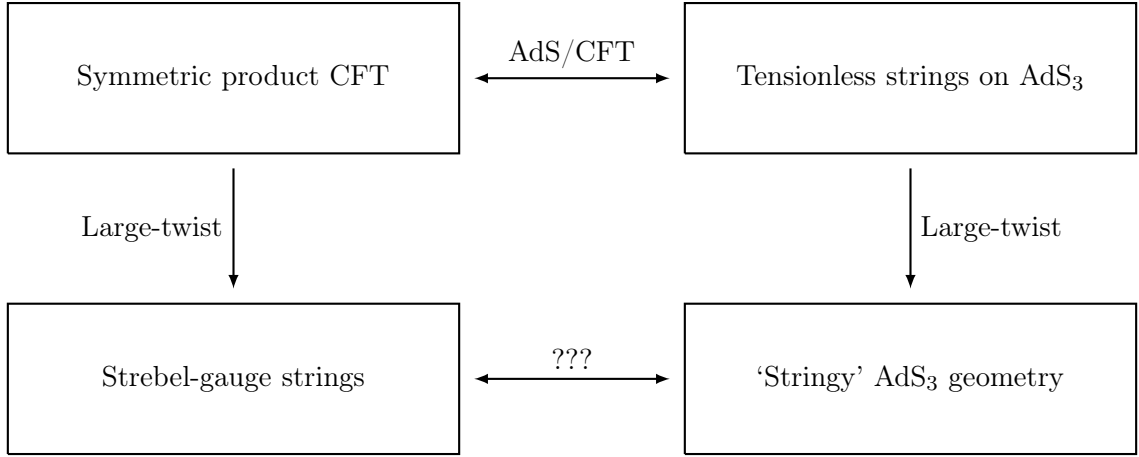

\vspace{0.5cm}

\begin{figure}[!ht]
\centering
\begin{tikzpicture}
\begin{scope}[xshift = -4.75cm, rotate = 30]
\draw[thick] (0,0) circle (3);
\path[draw, closed=true, thick, black, opacity = 0.7] (3,0) to[out = 135, in = -45] (2.25,1.5) to[out = 135, in = -45] (0.2,2.98) to[out = -135, in = 45] (-2,1.8) to[out = -135, in = 45] (-3,0) to[out = -45, in = 135] (-1.7,-2) to[out = -45, in = 135]  (-0.2,-2.98) to[out = 45, in = -135] (2,-1.8) to[out = 45, in = -135] (3,0);
\path[fill, closed=true, thick, blue, opacity = 0.05] (3,0) to[out = 135, in = -45] (2.25,1.5) to[out = 135, in = -45] (0.2,2.98) to[out = -135, in = 45] (-2,1.8) to[out = -135, in = 45] (-3,0) to[out = -45, in = 135] (-1.7,-2) to[out = -45, in = 135]  (-0.2,-2.98) to[out = 45, in = -135] (2,-1.8) to[out = 45, in = -135] (3,0);
\path[draw, closed=true, thick, black, opacity = 0.7] (3,0) to[out = 135, in = -45] (2,2) to[out = 135, in = -45] (0.2,2.98) to[out = -135, in = 45] (-2,2) to[out = -135, in = 45] (-3,0) to[out = -45, in = 135] (-2,-2) to[out = -45, in = 135]  (-0.2,-2.98) to[out = 45, in = -135] (2,-2) to[out = 45, in = -135] (3,0);
\path[fill, closed=true, thick, blue, opacity = 0.05] (3,0) to[out = 135, in = -45] (2,2) to[out = 135, in = -45] (0.2,2.98) to[out = -135, in = 45] (-2,2) to[out = -135, in = 45] (-3,0) to[out = -45, in = 135] (-2,-2) to[out = -45, in = 135]  (-0.2,-2.98) to[out = 45, in = -135] (2,-2) to[out = 45, in = -135] (3,0);
\path[draw, closed=true, thick, black, opacity = 0.7] (3,0) to[out = 135, in = -45] (1.5,2.25) to[out = 135, in = -45] (0.2,2.98) to[out = -135, in = 45] (-1.8,2) to[out = -135, in = 45] (-3,0) to[out = -45, in = 135] (-2,-1.7) to[out = -45, in = 135]  (-0.2,-2.98) to[out = 45, in = -135] (1.8,-2) to[out = 45, in = -135] (3,0);
\path[fill, closed=true, thick, blue, opacity = 0.05] (3,0) to[out = 135, in = -45] (1.5,2.25) to[out = 135, in = -45] (0.2,2.98) to[out = -135, in = 45] (-1.8,2) to[out = -135, in = 45] (-3,0) to[out = -45, in = 135] (-2,-1.7) to[out = -45, in = 135]  (-0.2,-2.98) to[out = 45, in = -135] (1.8,-2) to[out = 45, in = -135] (3,0);
\end{scope}
\draw[thick, -latex] (-1,0) -- (1,0);
\node[above] at (0,0) {$\text{deg}(\Gamma)\to\infty$};
\begin{scope}[xshift = 4.75cm, scale = 0.75, rotate = -15]
\fill[blue, opacity = 0.1] (0,0) circle (4);
\fill[white] plot[smooth, tension = 0.5] coordinates {(2.829,2.831) (0,0.6) (-2.829,2.831)} -- (0,5);
\fill[white] plot[smooth, tension = 0.5] coordinates {(2.831,2.829) (0.6,0) (2.831,-2.829)} -- (5,0);
\fill[white] plot[smooth, tension = 0.5] coordinates {(2.829,-2.831) (0,-0.6) (-2.829,-2.831)} -- (0,-5);
\fill[white] plot[smooth, tension = 0.5] coordinates {(-2.831,-2.829) (-0.6,0) (-2.831,2.829)} -- (-5,0);
\draw[thick] (0,0) circle (4);
\draw[thick] plot[smooth, tension = 0.5] coordinates {(2.829,2.831) (0,0.6) (-2.829,2.831)};
\draw[thick] plot[smooth, tension = 0.5] coordinates {(2.831,2.829) (0.6,0) (2.831,-2.829)};
\draw[thick] plot[smooth, tension = 0.5] coordinates {(2.829,-2.831) (0,-0.6) (-2.829,-2.831)};
\draw[thick] plot[smooth, tension = 0.5] coordinates {(-2.831,-2.829) (-0.6,0) (-2.831,2.829)};
\end{scope}
\end{tikzpicture}
\caption{Left: The worldsheet `glued' to the $\text{AdS}_3$ boundary for finite $N=\text{deg}(\Gamma)$. Right: The semiclassical worldsheet corresponding to the Strebel-gauge metric $g_{z\bar{z}}=N^2|\varphi|/4$ in the limit $N\to\infty$. The worldsheet near $z=z_i$ is approximated by a semi-infinite tube situated radially in $\text{AdS}_3$.}
\label{fig:tubes}
\end{figure}
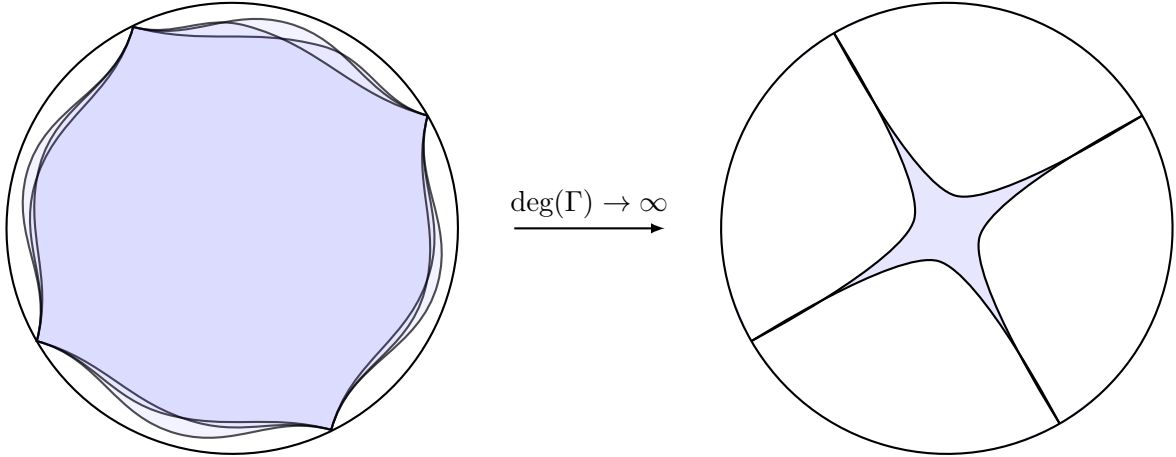

While the two above approaches are both formulated in the limit in which the worldsheet covers the $\text{AdS}_3$ boundary many times, they do not seem to have much to do with each other. In the first half of this paper, we explore the relationship between the two above observations, attempting to build a bridge between the large-twist limit of the worldsheet in terms of bulk geometry and the large-twist limit of the symmetric orbifold theory in terms of the Strebel differential (see Figure \ref{fig:approaches}). We will show, in particular, that when the degree of the covering map $\Gamma:\Sigma\to\mathbb{CP}^1$ describing the motion of the string is large, the string worldsheet no longer lives entirely at the boundary of $\text{AdS}_3$, but instead moves radially inward toward the bulk (see Figure \ref{fig:tubes}). In this limit, the induced metric on the worldsheet, which is given by the pullback of the $\text{AdS}_3$ metric, reproduces precisely the Nambu-Goto action \eqref{eq:intro-ng}, thus making explicit the geometric nature of the expression \eqref{eq:strebel-gauge-NG}. We also explore the worldsheet contributions to the partition functions in thermal $\text{AdS}_3$, which are represented by unramified coverings of the boundary torus by the worldsheet torus \cite{Eberhardt:2018,Eberhardt:2020bgq}. In this case, when the degree of the covering becomes large, the string worldsheet contracts to one which lives entirely near the center of $\text{AdS}_3$. The string then effectively becomes a massive particle moving in the bulk of $\text{AdS}_3$, thus giving a physical explanation for the appearance of conical defects in the stringy geometry of \cite{Eberhardt:2021jvj}.

In the second half of the paper, we explore a bit more the consequences of the relationship between the large-twist limit and the worldsheet theory. One of the hallmarks of the tensionless string is the realisation of the worldsheet theory in terms of a `twistorial' free field theory on the worldsheet \cite{Eberhardt:2018,Dei:2020}. These worldsheet free fields obtain classical configurations `on-shell', i.e. on the locus for which a covering map $\Gamma:\Sigma\to\mathbb{CP}^1$ exists. We show that these classical configurations for the twistorial fields are expressible as the solutions of a simple differential equation which, in the large-twist limit, relates the twistor fields to the Strebel differential in a surprisingly simple way. This provides a hint that the symmetric orbifold theory `knows' about the twistorial construction of the tensionless string. Armed with this understanding, we comment on the implications of the holographic duals of generic free CFTs.

As another application of the large-twist limit, we consider the motion of strings in an $\text{AdS}_3$ background in the presence of a totally reflective D1 brane localised in time, which thus carries the geometry of euclidean $\text{AdS}_2$. By analysing the effective action coming from the Strebel differential in the large-twist limit, we find that, in the limit that we ignore the directions transverse to the D1 brane, the endpoints of the string are described by an effective action which formally resembles the one-dimensional Schwarzian action describing boundary gravitons in Jackiw-Teitelboim (JT) gravity. We thus conjecture a relationship between the tensionless string and JT gravity in this compactification limit.

The outline of this paper is as follows: in Section \ref{sec:large-twist}, we review the construction of \cite{Gaberdiel:2020ycd} of correlation functions of the symmetric orbifold in the large-twist limit, emphasising how \eqref{eq:strebel-gauge-NG} arises from a sum over covering maps, and how the Strebel differential naturally emerges. We also exemplify this analysis by examining the covering maps which contribute to the (connected) contribution of the genus-one partition function of the symmetric orbifold. In Section \ref{sec:semiclassics}, we discuss the motion of a classical string in $\text{AdS}_3$ in the Wakimoto representation, and show that the large-twist limit classically describes strings which are allowed to move radially and thus explore the bulk. In doing so, we naturally recover the Strebel-gauge metric \eqref{eq:intro-ng} from the semiclassical worldsheet description. In Section \ref{sec:reconstruction}, we relate the semiclassical description of the worldsheet in terms of Strebel differentials to the twistorial worldsheet theory of \cite{Eberhardt:2018,Dei:2020}, and in particular show that the twistorial fields arise naturally as solutions to a holomorphic Schr\"odinger equation with the Strebel differential as the potential. In Section \ref{sec:ads2}, we extend our analysis to a background with a totally reflective brane with Euclidean $\text{AdS}_2$ geometry, and show that, in a certain compactification limit, the dynamics of the string worldsheet are described by the one-dimensional Schwarzian field theory of JT gravity. In Section \ref{sec:discussion}, we summarise our findings and discuss future directions. Finally, in Appendix \ref{sec:free-fields}, we review the free field construction of \cite{Eberhardt:2018,Dei:2020}, and explore a certain `twist' of the worldsheet theory which more readily meshes in with the analysis of Section \ref{sec:reconstruction} and for which correlation functions can be defined without the $W$ fields of \cite{Dei:2020}.

\section{The large-twist limit of the symmetric orbifold}\label{sec:large-twist}

In this section we review the analysis of \cite{Gaberdiel:2020ycd} in the large-twist limit of symmetric orbifold correlators. While aiming to be self-contained, we will leave out many details. In particular, we will not discuss the diagrammatic interpretation of the large-twist limit.

\subsection{The scattering equations}

As we mentioned in the introduction, if we consider states $\mathcal{O}_i^{(w_i)}$ in the symmetric orbifold theory, labelled by some twist $w_i$ and auxiliary index $i$ in the seed theory $\mathcal{M}$, we can consider the correlator of the form
\begin{equation}
\Braket{\prod_{i=1}^{n}\mathcal{O}_{i}^{(w_i)}(x_i)}\,.
\end{equation}
Here, we will consider the theory formulated on the sphere $\mathbb{CP}^1$. Such correlation functions enjoy a diagrammatic expansion in terms of covering spaces $\Gamma:\Sigma\to\mathbb{CP}^1$, and the `connected' component can be expressed as a sum over such maps \cite{Lunin_2001,Hamidi:1986vh}, weighted by an appropriate value of the string coupling and a conformal anomaly
\begin{equation}
\Braket{\prod_{i=1}^{n}\mathcal{O}_i^{(w_i)}(x_i)}_{\text{c}}=\sum_{\Gamma:\Sigma\to\mathbb{CP}^1}g_s^{2g-2+n}e^{-S_{\text{L}}[\Phi_\Gamma]}\Braket{\prod_{i=1}^{n}\mathcal{O}_i(z_i)}_{\Sigma}\,,
\end{equation}
where $g$ is the genus of the covering surface and $z_i$ are the critical points of $\Gamma$ on $\Sigma$, i.e.
\begin{equation}
\Gamma(z)=x_i+\mathcal{O}((z-z_i)^{w_i})\,,\quad z\to z_i\,.
\end{equation}
Here, $S_{\text{L}}$ is the Liouville action on the covering surface $\Sigma$
\begin{equation}\label{eq:liouville-covering}
S_{\text{L}}[\Phi]=\frac{c}{48\pi}\int_{\Sigma}(2\partial\Phi\,\bar{\partial}\Phi+R\,\Phi)\,,
\end{equation}
and $\Phi_{\Gamma}$ is the `classical' Liouville field associated to the covering map $\Gamma$
\begin{equation}
\Phi_{\Gamma}=\log|\partial\Gamma|^2\,.
\end{equation}

The main point to note is that all of the information about the correlation functions of twisted correlators in the symmetric orbifold is contained in the correlator of the seed theory on arbitrary surfaces $\Sigma$ and in the holomorphic data of the covering map $\Gamma$. Thus, assuming the seed theory is under sufficiently good control, we can reduce the problem of computing correlators in the symmetric orbifold to the problem of constructing a covering map with the desired analytic behaviour.\footnote{Strictly speaking, the Liouville action defined above diverges for $\Phi=\Phi_{\Gamma}$, and so one also needs to introduce a regularisation scheme, as in \cite{Lunin_2001}.}

Generically, given the data $\{x_i\}$, $\Sigma\setminus\{z_i\}$ and $\{w_i\}$, finding such a covering map is an algebraically difficult problem, and can only be solved when the moduli of $\Sigma\setminus\{z_i\}$ are finely-tuned. In fact, given points on the sphere $\{x_i\}$ and twists $\{w_i\}$, there are only finitely many points in the moduli space $\mathcal{M}_{g,n}$ for which a covering map $\Gamma:\Sigma\setminus\{z_i\}\to\mathbb{CP}^1\setminus\{x_i\}$ exists with the correct critical behaviour near $z_i$.

For now, let us restrict to the case where our covering surface has genus zero, so that we are considering branched covering maps $\Gamma:\mathbb{CP}^1\to\mathbb{CP}^1$. In this case, the degree of $\Gamma$ can be determined by the Riemann-Hurwitz formula as
\begin{equation}
\text{deg}(\Gamma)=N=1+\sum_{i=1}^{n}\frac{w_i-1}{2}\,.
\end{equation}
Thus, since $\Gamma$ is a holomorphic function on $\mathbb{CP}^1$ (or equivalently a meromorphic function on $\mathbb{C}$), we can write
\begin{equation}
\Gamma(z)=\frac{Q_N(z)}{P_N(z)}
\end{equation}
for polynomials $Q_N,P_N$ of degree $N$.\footnote{We are assuming none of the $x_i$ and none of the $z_i$ lie at infinity.} Let us call the zeroes of $P_N$ $\lambda_a$ for $a=1,\ldots,N$.

Now, let us consider the derivative $\partial\Gamma$, which will be a ratio of polynomials of degree $2N$. It is easy to see that $\Gamma$ must be of the form
\begin{equation}
\partial\Gamma(z)=C\frac{\prod_{i=1}^{n}(z-z_i)^{w_i-1}}{\prod_{a=1}^{N}(z-\lambda_a)^2}\,,
\end{equation}
since $\partial\Gamma$ has zeroes of order $w_i-1$ at $z=z_i$ and poles of order 2 at $z=\lambda_a$. The numerator has degree $2N-2$ so that $\partial\Gamma(z)\sim 1/z^2$ as $z\to\infty$. Now, since $\partial\Gamma$ is a total derivative of a meromorphic function, its residue at any pole must vanish, since otherwise $\Gamma$ would have logarithmic contributions to its Laurent series expansion around $z=\lambda_a$. Demanding that the residue at $\lambda_a$ vanishes leads to the so-called `scattering equations' \cite{Roumpedakis:2018tdb}
\begin{equation}\label{eq:scattering-equations}
\sum_{i=1}^{n}\frac{w_i-1}{\lambda_a-z_i}=\sum_{b\neq a}\frac{2}{\lambda_a-\lambda_b}\,,
\end{equation}
which are the central algebraic constraints for the existence of a covering map (of course, one also has to demand that $\Gamma(z_i)=x_i$ for all $i$ once the scattering equations are solved).

\subsection{The matrix model and the spectral curve}

The insight of \cite{Gaberdiel:2020ycd} was to rewrite \eqref{eq:scattering-equations} in terms of the classical equations of motion of an $N\times N$ Hermitian matrix model. Indeed, consider the matrix integral
\begin{equation}
\int\mathrm{d}M\,\exp\left(-V(M)\right)\,,
\end{equation}
where the potential is a so-called Penner-like potential
\begin{equation}
V(M)=N\sum_{i=1}^{n}\alpha_i\log(M-z_i\textbf{1})\,.
\end{equation}
We can now diagonalise $M$ in terms of eigenvalues $\lambda_a$, and recast the matrix integral into the form
\begin{equation}
\int\mathrm{d}^N\lambda\,\Delta(\lambda)^2\,\exp\left(-N\sum_{a=1}^{N}\sum_{i=1}^{n}\alpha_i\log(\lambda_a-z_i)\right)\,,
\end{equation}
where $\Delta(\lambda)^2=\prod_{a<b}(\lambda_a-\lambda_b)^2$ is the usual Vandermonde determinant arising upon diagonalisation. The inclusion of the Vandermode determinant induces an effective potential given by
\begin{equation}
V_{\text{eff}}(z)=\sum_{i=1}^{n}\alpha_i\log(z-z_i)-\frac{2}{N}\sum_{a=1}^{N}\log(z-\lambda_a)\,,
\end{equation}
such that the partition function of the matrix model can be written as\footnote{One needs to be careful with the term $\log(\lambda_a-\lambda_a)=-\infty$ in the effective potential; however, this only introduces an (infinite) constant, and thus does not affect the dynamics of the theory.}
\begin{equation}
\int\mathrm{d}^N\lambda\,\exp\left(-N\sum_{a=1}^{N}V_{\text{eff}}(\lambda_a)\right)\,.
\end{equation}

The matrix model integral is generally computed by first considering the tree-level contributions, i.e. the configurations such that $V_{\text{eff}}'(\lambda_a)=0$ for all $a$ (and indeed for $N\gg 1$, these are the dominant contributions). These classical equations of motion are
\begin{equation}
\sum_{i=1}^{n}\frac{\alpha_i}{\lambda_a-z_i}=\frac{2}{N}\sum_{b\neq a}\frac{1}{\lambda_a-\lambda_b}\,.
\end{equation}
Thus, we recover precisely the scattering equations \eqref{eq:scattering-equations} with $\alpha_i=(w_i-1)/N$.\footnote{Strictly speaking, the solutions $\lambda_a$ to the scattering equations that we are interested in are complex, as opposed to real solutions one would expect from a Hermitian matrix model. Since the matrix model defined above is only used as an auxilliary object, this subtlety causes no harm, and we can simply take $z_i,\lambda_a\in\mathbb{C}$.}

For $N\gg 1$, there are powerful methods for computing the solutions to the saddle-point equations for matrix models which we now have at our disposal. In particular, we assume that at large $N$ the eigenvalues $\lambda_a$ condense into a curve in the complex plane, which potentially has many disconnected components. Let $\mathcal{C}=\bigcup_{\ell=1}^{m}\mathcal{C}_\ell$ be the decomposition of this curve into disjoint segments. Then we can write the scattering equations as
\begin{equation}\label{eq:riemann-hilbert}
\sum_{i=1}^{n}\frac{\alpha_i}{\lambda-z_i}=2P\int_{\mathcal{C}}\frac{\rho(\lambda')\,\mathrm{d}\lambda'}{\lambda-\lambda'}\,,
\end{equation}
where $\rho(\lambda)$ is the asymptotic density of eigenvalues along the curve $\mathcal{C}$ and $P$ denotes the principle value of the integral.

Equation \eqref{eq:riemann-hilbert} defines a Riemann-Hilbert problem, whose solution can be found with the help of a so-called spectral curve. We define an auxiliary function
\begin{equation}\label{eq:y-definition}
y(z)=\sum_{i=1}^{n}\frac{\alpha_i}{z-z_i}-2\int_{\mathcal{C}}\frac{\rho(\lambda)\,\mathrm{d}\lambda}{z-\lambda}
\end{equation}
which is holomorphic and globally defined on $\mathbb{C}\setminus\mathcal{C}$. By the scattering equations \eqref{eq:riemann-hilbert}, we see that $y(z)=0$ on the endpoints of $\mathcal{C}$. Furthermore, $y$ has square-root branch cuts along $\mathcal{C}$, and is thus not globally defined on $\mathbb{C}$. However, its square $\phi(z)=y(z)^2$ is globally defined on $\mathbb{C}$. We can therefore think of $y$ as being globally defined on a Riemann surface $\tilde{\Sigma}$ defined via the polynomial equation
\begin{equation}
\tilde{\Sigma}:y^2=\phi(z)\,.
\end{equation}
The curve $\tilde{\Sigma}$ is known as the \textit{spectral curve} of the matrix model, and is a double cover of the Riemann sphere ramified at the points $\partial\mathcal{C}$. A natural basis of the homology group $H_1(\tilde{\Sigma},\mathbb{Z})$ is one in which cycles which surround branch cuts form the $A$-cycles of $\tilde{\Sigma}$, while the remaining cycles form the $B$-cycles. If $\mathcal{C}$ has $2m$ endpoints, then the genus of $\tilde{\Sigma}$ is $m-1$, since only $m-1$ of the $A$-cycles/$B$-cycles are independent.

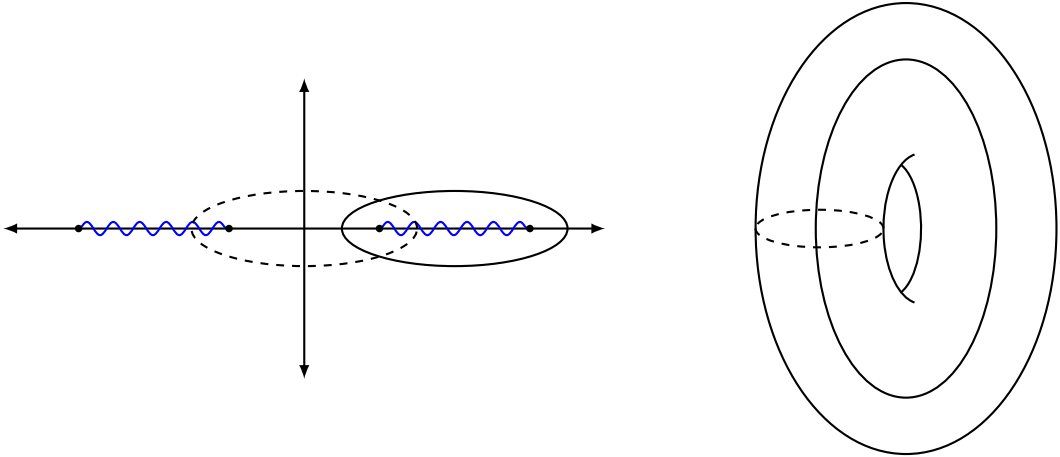
\begin{figure}
\centering
\begin{tikzpicture}
\begin{scope}[xshift = -4cm]
\draw[thick, latex-latex] (-4,0) -- (4,0);
\draw[thick, latex-latex] (0,2) -- (0,-2);
\draw[thick, snake it, blue] (-3,0) -- (-1,0);
\fill (-3,0) circle (0.05);
\fill (-1,0) circle (0.05);
\draw[thick, snake it, blue] (1,0) -- (3,0);
\fill (1,0) circle (0.05);
\fill (3,0) circle (0.05);
\draw[thick] (2,0) [partial ellipse = 0:360:1.5 and 0.5];
\draw[thick, dashed] (0,0) [partial ellipse = 0:360:1.5 and 0.5];
\end{scope}
\begin{scope}[xshift = 4cm]
\draw[thick] (0,0) [partial ellipse = 0:360:2 and 3];
\draw[thick] (0.2,0) [partial ellipse = 100:260:0.5 and 1];
\draw[thick] (-0.2,0) [partial ellipse = -70:70:0.4 and 0.9];
\draw[thick] [partial ellipse = 0:360:1.2 and 2.25];
\draw[thick, dashed] (-2+0.85,0) [partial ellipse = 180:360:0.85 and 0.25];
\draw[thick, dashed] (-2+0.85,0) [partial ellipse = 0:180:0.85 and 0.25];
\end{scope}
\end{tikzpicture}
\caption{Left: Non-contractible cycles of a genus-1 spectral curve $\tilde{\Sigma}:y^2=\phi(z)$. Squiggly blue lines represent the branch cuts, while solid cycles represent the $A$-cycles of $\tilde{\Sigma}$ and dashed cycles represent the $B$-cycles. Right: The lift of the non-contractible cycles onto the genus-1 covering surface.}
\end{figure}

Once the spectral curve is known, it can be used to reconstruct the density $\rho(\lambda)$, and thus the solution of the scattering equations. Indeed, since $\mathcal{C}$ is essentially just the branch cut of a function $y=\sqrt{\phi}$, it can be chosen arbitrarily such that its endpoints are the zeroes of the function $y$. Once we know $y$, we can then determine the density $\rho(\lambda)$ via the Sokhotski–Plemelj theorem, i.e. by taking the difference in the value of $y$ just above/below a branch cut
\begin{equation}\label{eq:density-from-y}
\rho(\lambda)=\frac{1}{4\pi i}(y(\lambda+i\varepsilon)-y(\lambda-i\varepsilon))\,,
\end{equation}
for $\lambda\in\mathcal{C}$. Furthermore, we can define the so-called \textit{filling fractions} of the density $\rho$ via periods of the differential $\sqrt{\phi}$. That is,
\begin{equation}
c_\ell=\int_{\mathcal{C}_{\ell}}\rho(\lambda)\,\mathrm{d}\lambda=\frac{1}{2\pi i}\oint_{\mathcal{C}_{\ell}}\sqrt{\phi(z)}\,\mathrm{d}z\,,
\end{equation}
where the integral domain of the second integral is the $A$-cycle of $\tilde{\Sigma}$ enclosing the branch cut $\mathcal{C}_\ell$. In terms of the covering map $\Gamma$, the filling fractions determine what fraction of the poles condense on to the component $\mathcal{C}_{\ell}$ of the curve $\mathcal{C}$.

\subsection{The Strebel differential}

The differential $\phi(z)\,(\mathrm{d}z)^2$ defining the spectral curve above has the following properties:
\begin{itemize}

	\item By \eqref{eq:y-definition}, near the insertion points $z=z_i$, we have
	\begin{equation}
	\phi(z)(\mathrm{d}z)^2\sim\frac{\alpha_i^2(\mathrm{d}z)^2}{(z-z_i)^2}+\cdots\,.
	\end{equation}

	\item $\phi$ has simple zeroes precisely at the endpoints of the branch-cuts $\mathcal{C}_{\ell}$ (let us call them $a_{\ell},b_{\ell}$, with $\partial\mathcal{C}_{\ell}=\{a_{\ell},b_{\ell}\}$).

	\item If $\gamma$ is a path connecting the zeroes $a_{\ell}$ and $b_{\ell}$, then
	\begin{equation}
	\frac{1}{2\pi i}\int_{\gamma}\sqrt{\phi(z)}\,\mathrm{d}z\in\mathbb{R}\,.
	\end{equation}

\end{itemize}
This last point is due to \eqref{eq:density-from-y}, since $\rho(\lambda)$ is taken to be real. These properties identity what is known in the mathematical literature as a \textit{Strebel differential} (see Appendix A of \cite{Maity:2021wsi} for a gentle introduction). Let us define a quadratic differential on $\Sigma$ via
\begin{equation}\label{eq:strebel-differential}
\varphi(z)=-\phi(z)\,(\mathrm{d}z)^2\,.
\end{equation}
It is a theorem of Strebel that given values $\alpha_i\in\mathbb{R}^+$, there exists a unique quadratic differential $\varphi$ satisfying the above properties for any given value of the moduli $\{z_i\}$. Thus, for each choice of the moduli $\{z_i\}$, we find a unique spectral curve, and thus we can uniquely solve the scattering equations \eqref{eq:riemann-hilbert}.

Let us now return to the problem of constructing the covering map. Note that for finite $N$, we have
\begin{equation}
\frac{1}{N}\log\partial\Gamma(z)=\sum_{i=1}^{n}\alpha_i\log(z-z_i)-\frac{2}{N}\sum_{a=1}^{N}\log(z-\lambda_a)+\frac{1}{N}\log C
\end{equation}
which is the effective potential of our matrix model, and we have ignored a constant contribution. Taking $N$ large, we can write
\begin{equation}
\frac{1}{N}\log\partial\Gamma(z)=\sum_{i=1}^{n}\alpha_i\log(z-z_i)-2\int_{\mathcal{C}}\mathrm{d}\lambda\,\rho(\lambda)\log(z-\lambda)\,.
\end{equation}
Finally, we can take the derivative and we find
\begin{equation}
\frac{1}{N}\frac{\partial^2\Gamma}{\partial\Gamma}=\sum_{i=1}^{n}\frac{\alpha_i}{z-z_i}-2\int_{\mathcal{C}}\frac{\rho(\lambda)\,\mathrm{d}\lambda}{z-\lambda}=y(z)\,.
\end{equation}
The branch cuts of $y$ can be thought of as arising from the coalescence of poles of $\Gamma$ along the curve $\mathcal{C}$.\footnote{For a simple example of this phenomenon, consider a function $f$ defined by
\begin{equation}
f(z)=\frac{1}{N}\sum_{a=1}^{N}\frac{1}{z-a/N}\,.
\end{equation} As $N\to\infty$, we have $f(z)\to\log(1-z)+\log(z)$, which has a branch cut along $[0,1]$.} Thus, we can relate the spectral curve coordinate $y$ directly to the covering map. This allows us to immediately write down the Strebel differential \eqref{eq:strebel-differential} as
\begin{equation}\label{eq:strebel-covering}
\varphi=-\frac{1}{N^2}\left(\frac{\partial^2\Gamma}{\partial\Gamma}\right)^2\,.
\end{equation}

Finally, as noted in \cite{Gaberdiel:2020ycd}, at large $N$ we can approximate $\varphi$ via the Schwarzian derivative of $\Gamma$, i.e.
\begin{equation}
\varphi=\frac{2}{N^2}S[\Gamma]=\frac{2}{N^2}\left(\partial\left(\frac{\partial^2\Gamma}{\partial\Gamma}\right)-\frac{1}{2}\left(\frac{\partial^2\Gamma}{\partial\Gamma}\right)^2\right)\,.
\end{equation}
In the large $N$ limit, the first term in the Schwarzian is subdominant, and we are simply left with \eqref{eq:strebel-covering}. As we will see later, however, writing the Strebel differential in terms of the Schwarzian derivative of the covering map allows for a natural interpretation of the $\text{AdS}_3$ worldsheet theory.

\subsection{The Strebel-gauge action}

Now let us return to the problem of constructing correlation functions of the symmetric orbifold, which is computed in terms of the Liouville action of the field $\Phi_{\Gamma}=\log|\partial\Gamma|^2$. Note that one can immediately write the derivative of the Liouville field in terms of the spectral curve $y(z)$, i.e.
\begin{equation}
\partial\Phi_{\Gamma}=Ny\,,\quad\bar{\partial}\Phi_{\Gamma}=N\bar{y}\,,
\end{equation}
and thus the Liouville action can be expressed as
\begin{equation}\label{eq:Strebel-Liouville}
S_{L}[\Phi_{\Gamma}]=\frac{cN^2}{24\pi}\int_{\Sigma}|y|^2=\frac{N^2}{4\pi}\int_{\Sigma}|\varphi|\,,
\end{equation}
where $\varphi$ is the Strebel differential, and in the second equality we have specified to the $\mathbb{T}^4$ (or $\text{K}3$) seed theory, i.e. $c=6$. If we are working on a covering surface with genus $g$ and $n$ punctures, there is a unique Strebel differential for each value of $\alpha_i$ and each value of the moduli of the covering space. Thus, we can express the symmetric product orbifold correlation functions in this limit as an integral over $\mathcal{M}_{g,n}$, namely
\begin{equation}
\Braket{\prod_{i=1}^{n}\mathcal{O}_{i}^{w_i}(x_i)}_{\text{c}}\sim\sum_{g=0}^{\infty}g_s^{n+2g-2}\int_{\mathcal{M}_{g,n}}\exp\left(-\frac{N^2}{4\pi}\int_{\Sigma}|\varphi|\right)\Braket{\prod_{i=1}^{n}\mathcal{O}_i(z_i)}_{\Sigma}\,.
\end{equation}

The effective action \eqref{eq:Strebel-Liouville} and the integral over $\mathcal{M}_{g,n}$ has an immediate stringy interpretation. Given a Strebel differential $\varphi$, one can associate a canonical metric $g_{z\bar{z}}=|\varphi|/4$ such that $g$ is flat except at the poles and zeroes of $\varphi$, which give curvature singularities of opposite sign. The action \eqref{eq:Strebel-Liouville} then has the interpretation of the Nambu-Goto action of some minimal area string whose pullback metric is $g$ and whose tension is $T=N^2$. Since this action arose naturally from a 2D CFT with known $\text{AdS}_3$ dual, one would expect that $g$ is the pullback of the $\text{AdS}_3$ metric on the worldsheet. Indeed, in Section \ref{sec:semiclassics} we will see that this is the case precisely when one takes $N$ to be large.

\subsection{Example: the partition function}\label{sec:partition-function}

In the above subsections, we reviewed the construction of the Strebel differential in the large-twist limit of the symmetric orbifold. In order to exemplify how this might work, let us consider the correlation function for which the covering map is as simple as possible: the genus-one partition function. Here, $\Gamma$ is an unramified covering map between two tori, the worldsheet and the boundary of thermal $\text{AdS}_3$, and its degree can be taken to be as large as possible without introducing analytic difficulties.

To compute the partition function, we follow the approach of \cite{Bantay:1999us}. In order to keep track of the twist of the CFT states, we consider the `grand canonical' partition function
\begin{equation}
\mathfrak{Z}(p,t)=\sum_{K=0}^{\infty}p^KZ_{\text{Sym}^K(\mathcal{M})}(t)\,,
\end{equation}
where $p$ is a chemical potential counting the order $K$, and $t$ is the torus modulus. A fundamental result in the theory of permutation orbifolds is that this partition function can be expressed in terms of Hecke operators
\begin{equation}
\mathfrak{Z}(p,t)=\exp\left(\sum_{N=1}^{\infty}p^NT_NZ(t)\right)\,,
\end{equation}
where $Z(t)$ is the partition function of the seed theory $\mathcal{M}$, and the Hecke operators $T_N$ are defined below. Intuitively, the term in the exponential is generated by connected covering maps of degree $w$, which are always tori. The part of this partition function which is visible from the worldsheet theory is the connected component $\log\mathcal{Z}(p,t)$, and the full partition function can only be recovered by considering a second quantised theory of strings.

Let us examine the form of the connected component of the partition function at large $N$. Recall that the Hecke operator $T_N$ acts on modular functions as
\begin{equation}\label{eq:Hecke-operator}
T_NZ(t)=\frac{1}{N}\sum_{ad=N}\sum_{b=0}^{d-1}Z\left(\frac{at+b}{d}\right)\,.
\end{equation}
The modulus $\tau=(at+b)/d$ is to be interpreted as the modulus of a torus which covers the original torus with degree $N$ and is therefore interpreted as the modulus of the dual worldsheet. The number of such covering tori is given by the \textit{divisor function} of $N$
\begin{equation}
\sum_{ad=N}\sum_{b=0}^{d-1}1=\sum_{d|N}d=\sigma_1(N)\,.
\end{equation}
The growth rate of this function is known to be asymptotically bounded by as $\mathcal{O}(N\log\log N)$ \cite{Hardy:1938}, and so the Hecke operator diverges at worst double logarithmically as $N\to\infty$.\footnote{In fact, assuming the Riemann hypothesis is true, it can be shown that $\sigma_1(N)<e^{\gamma}N\log\log N$ \cite{Ramanujan:1997}. If the reader is uncomfortable with convergence issues, we can simply take $N$ prime, for which $\sigma_1(N)=N+1$, and the Hecke operator converges, assuming $Z(\tau)$ is reasonably well-behaved as $\tau\to i\infty$.}

\begin{figure}
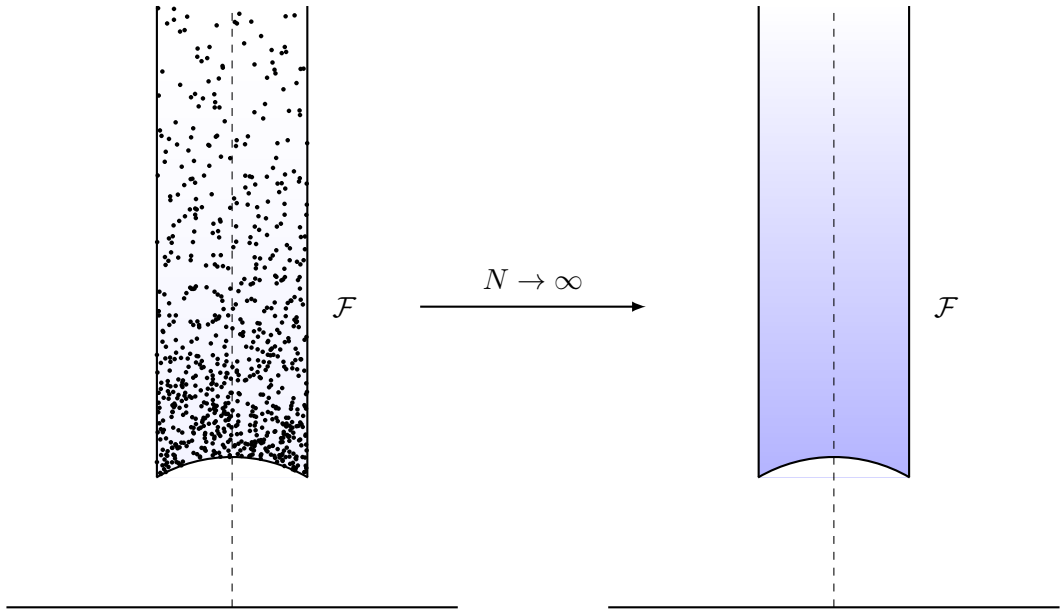

\centering

\caption{Left: the distribution of covering tori for $N=971$ and $t=0.2+2i$. Moduli with $\text{Im}(\tau)>4$ have been omitted. Right: the limiting probability distribution as $N\to\infty$.}
\label{fig:moduli-distribution}
\end{figure}

We can thus, naively, consider the limit in which $N\to\infty$, for which the number of covering spaces diverges, and it seems reasonable that the space of allowed $\tau$ forms a continuum. To exemplify how this works, we can consider the simple case of $N$ prime. Since $N$ only has two divisors, $1$ and itself, the moduli of the covering tori can be written as
\begin{equation}\label{eq:moduli-list}
\left\{Nt,\frac{t}{N},\ldots,\frac{t+N-1}{N}\right\}\,.
\end{equation}
As $N\to\infty$, we can forget the first element of \eqref{eq:moduli-list} without affecting the sum too much, and focus simply on the moduli of the form $\tau=(t+b)/N$. However, these moduli do not all lie in the fundamental domain, and will generically all lie in different fundamental domains of $\text{SL}(2,\mathbb{Z})$. If we bring them into the fundamental domain, their distribution will be effectively random, since the modular transformations bringing each modulus into $\mathcal{F}$ differ wildly as we change $b$ (see Figure \ref{fig:moduli-distribution} for an example with $N=971$). More concretely, the map $\mathbb{H}\to\mathcal{F}$ taking a point in the upper-half-plane into its representative in the fundamental domain exhibits chaotic behaviour as $\text{Im}(\tau)\to 0$.

As we increase $N$ (but keeping $N$ prime), we can approximate the Hecke operator $T_N$ as the following integral
\begin{equation}
T_NZ(t)\approx\lim_{\varepsilon\to 0}\int_{0}^{1}\mathrm{d}x\,Z(x+i\varepsilon)\,,
\end{equation}
where we have defined the small number $\varepsilon=t/iN$. The line $x+i\varepsilon$ for $x\in[0,1)$ equidistributes into the fundamental domain $\mathcal{F}$ as $\varepsilon\to 0$ with measure $\mathrm{d}^2\tau/\text{Im}(\tau)^2$ \cite{mathoverflowthing}.\footnote{We thank Lorenz Eberhardt for pointing this out.} Thus, we have
\begin{equation}
T_NZ(t)\longrightarrow\frac{3}{\pi}\int_{\mathcal{F}}\frac{\mathrm{d}^2\tau}{\text{Im}(\tau)^2}\,Z(\tau)\,,
\end{equation}
where the prefactor is simply $\text{Vol}(\mathcal{F})^{-1}$, which we will ignore from now on. Therefore, the symmetric orbifold partition function for fixed $N$ approaches the path integral of a string propagating on the seed theory $\mathcal{M}$ as $N\to\infty$, which we will interpret below as the string propagating only through the origin of $\text{AdS}_3$ and only having fluctuations in the $\mathcal{M}$-direction.

In the language of the Strebel differential, we might try to immediately write down an answer as
\begin{equation}
\int_{\mathcal{F}}\frac{\mathrm{d}^2\tau}{\text{Im}(\tau)^2}\,Z(\tau)e^{-S_{\text{NG}}(\tau)}\,,
\end{equation}
where the measure $\mathrm{d}\mu$ is the same as above, since it arises from the density of covering maps in $\mathcal{F}$ as $N\to\infty$. The above analysis then seems to suggest that $S_{\text{NG}}(\tau)=0$ for all $\tau$. The reason for this is quite simple: for the configuration above, the Strebel differential vanishes.\footnote{This naively contradicts the statement that Strebel differential are in one-to-one correspondence to points in the moduli space $\mathcal{M}_{g,n}\times\mathbb{R}^n$. However, this statement only holds for surfaces with $n+2g-2>0$, i.e. for which the space of holomorphic quadratic differentials is non-vanishing. Mathematically, this corresponds to the stability condition of the moduli space $\mathcal{M}_{g,n}$.} Indeed, as we will argue in a moment, the covering map $\Gamma$ from the worldsheet torus to the AdS boundary torus is linear, and thus its corresponding Strebel differential vanishes.

The covering map from the worldsheet torus to the spacetime boundary is given by $\Gamma(z)=\alpha z$ for some constant $\alpha$. The periodicity conditions require $\Gamma(z+1)=\Gamma(z)+ct+d$ and $\Gamma(z+\tau)=\Gamma(z)+at+b$ for integers $a,b,c,d$. This requires $\alpha=ct+d$ and $\tau=(at+b)/(ct+d)$. As above, we can take $c=0$ and $0\leq b\leq d-1$. Thus, the covering map is $\Gamma(z)=dz$. The integer $d$ determines the number of times the worldsheet wraps the contractible cycle of $\text{AdS}_3$, and in general will be large in the sum.

From the point of view of the worldsheet, the vanishing of the Strebel action is quite surprising, since we expect it to reproduce the area of some semiclassical string worldsheet in $\text{AdS}_3$. In fact, in the limit in question, this area vanishes. As we will see in the next section, in the limit of large $d$ this covering map describes a string completely localised in the center of $\text{AdS}_3$, and thus has zero area (see Figure \ref{fig:massive-particle} below). The only part of the worldsheet theory where the string has proper dynamics, then, is the internal manifold $\mathcal{M}$, and so the string partition function is simply the integral of $Z(\tau)$ over the moduli space $\mathcal{M}_{1,0}$.\footnote{Strictly speaking, the string also propagates in the $\text{S}^3$. The motion of the string in this direction should be apparent if we consider the full supersymmetric generalization of this analysis.} We should also note that the string being localised to the center of $\text{AdS}_3$ supports the proposal of \cite{Eberhardt:2021jvj} that a string wrapping a contractible cycle in the bulk generates a conical defect, since a string localised to the center of $\text{AdS}_3$ simply behaves as a massive particle, which sources a defect singularity.

\section{Semiclassical analysis}\label{sec:semiclassics}

In the previous section we reviewed how the correlation functions of the symmetric product orbifold reorganise themselves into a path integral over string worldsheets whose areas are given by the Strebel metric $\mathrm{d}s^2=N^2|\varphi|\mathrm{d}z\,\mathrm{d}\bar{z}/4$. In this section, we explore the large-twist limit of a classical string sigma model on $\text{AdS}_3$ and show how the Strebel differential arises naturally in this context as the pullback of the $\text{AdS}_3$ metric onto the worldsheet, providing a link between the large-twist limit of the symmetric orbifold and that of the $\text{AdS}_3$ theory.

\subsection{The worldsheet sigma model}

Consider the semiclassical action of (bosonic) string theory on $\text{AdS}_3$, whose first-order form (see, e.g. \cite{Eberhardt:2019}) can be written as
\begin{equation}\label{eq:Wakimoto}
S=\frac{k}{4\pi}\int_{\Sigma}\mathrm{d}^2z\,\left(4\partial\Phi\,\bar{\partial}\Phi+\beta\bar{\partial}\gamma+\bar{\beta}\partial\bar{\gamma}-e^{-2\Phi}\beta\bar{\beta}-k^{-1}R\Phi\right)\,,
\end{equation}
where $R$ is the worldsheet curvature, and $k$ is the amount of NS-NS flux (the tensionless limit corresponds to $k=1$). Geometrically, the pair $(\gamma,\bar{\gamma})$ parametrises the motion of the string along the boundary of $\text{AdS}_3$ in complex coordinates, while the scalar $\Phi$ is related to the Poincar\'e radial coordinate as $r^2=e^{-2\Phi}$, so that $\Phi\to\infty$ is at the boundary of $\text{AdS}_3$.\footnote{Note that the normalisation of the radial coordinate $r$ is such that $r=0$ is the asymptotic boundary and $r=\infty$ is the center of AdS.} Given a holomorphic map $\Gamma:\Sigma\to\mathbb{CP}^1$, one can construct a solution to the classical equations of motion given by \cite{Eberhardt:2019}
\begin{equation}\label{eq:Wakimoto-classical}
\gamma(z,\bar{z})=\Gamma(z)\,,\quad\bar{\gamma}(z,\bar{z})=\bar{\Gamma}(\bar{z})\,,\quad\Phi(z,\bar{z})=\log\frac{1}{\varepsilon}-\frac{1}{2}\log|\partial\Gamma|^2\,.
\end{equation}
(There is a corresponding expression for $\beta$, but we will not need it). Here, $\varepsilon$ is a scaling parameter which can be thought of as an infrared cutoff. In the limit that $\varepsilon\to 0$, the corresponding classical solution satisfies $\Phi\to\infty$, and thus the worldsheet is `pinned' to the boundary. This is precisely the limit in which the action \eqref{eq:Wakimoto} becomes free, and so the path integral expanded around the solution \eqref{eq:Wakimoto-classical} becomes a Gaussian integral, and we can thus expect \eqref{eq:Wakimoto-classical} to be valid quantum mechanically, regardless of the value of $k$.

In the tensionless ($k=1$) string theory on $\text{AdS}_3\times\text{S}^3\times\mathcal{M}$, the classical solution can be shown to be the only contribution to the string path integral \cite{Eberhardt:2019,Dei:2020}. In particular, if we consider  vertex operators $V^{w_i}(x_i,z_i)$, where $w_i$ labels the spectral flow of the corresponding state, then the correlation function
\begin{equation}
\Braket{\prod_{i=1}^{n}V^{w_i}(x_i,z_i)}
\end{equation}
receives only contributions in the path integral from solutions of the form \eqref{eq:Wakimoto-classical}, where the holomorphic map $\Gamma$ satisfies
\begin{equation}\label{eq:covering-map}
\Gamma(z)\sim x_i+\mathcal{O}((z-z_i)^{w_i})
\end{equation}
near the worldsheet insertion points. Thus, taking the scaling limit $\varepsilon\to 0$, we are left with the conclusion that \textit{every} contribution to the string theory path integral is given by a string which is completely localised to the boundary of AdS. That is, for tensionless strings on $\text{AdS}_3\times\text{S}^3\times\mathcal{M}$, we can take seriously the expression \eqref{eq:Wakimoto-classical} as the motion of the worldsheet in the target space.

Naively, one would conclude, then, that the tensionless string theory on $\text{AdS}_3\times\text{S}^3\times\mathcal{M}$ knows nothing about the bulk, since for the solutions \eqref{eq:Wakimoto-classical} we have $\Phi\to\infty$ and so the string never probes the spacetime geometry beyond the boundary. For finite values of $w$, this is indeed true. However, note that if $w$ is taken to be very large (and, in particular, parametrically larger than the infrared cutoff $1/\varepsilon$), then we can actually consider worldsheets which probe the interior. To see this, let us expand the solution \eqref{eq:Wakimoto-classical} around the point $z=z_i$. We have
\begin{equation}
\Phi(z,\bar{z})=\log\frac{1}{\varepsilon w_i}-\frac{w_i-1}{2}\log|z-z_i|^2+\cdots
\end{equation}
The radial profile $r^2=e^{-2\Phi}$ then takes the form
\begin{equation}
r^2(z,\bar{z})=\varepsilon^2|\partial\Gamma|^2\sim\varepsilon^2w_i^2|z-z_i|^{w_i-1}\,.
\end{equation}
That is, if we take $w_i\gg 1/\varepsilon$, the radial profile can become non-zero, and the string is allowed to probe the bulk.

\begin{figure}
\centering
\begin{tikzpicture}
\begin{scope}[xshift = -4cm]
\fill[blue, opacity = 0.1] [partial ellipse = 0:360:1.7 and 2.7];
\fill[white] (-0.15,0) [partial ellipse = -80:80:0.7 and 1.2];
\fill[white] (0.25,0) [partial ellipse = 110.3:249.7:0.8 and 1.3];
\draw[thick] (0,0) [partial ellipse = 0:360:2 and 3];
\draw[thick] (0.2,0) [partial ellipse = 110:250:0.5 and 1];
\draw[thick] (-0.2,0) [partial ellipse = -70:70:0.4 and 0.9];
\draw[thick] (-2+0.85,0) [partial ellipse = 180:360:0.85 and 0.25];
\draw[thick, dashed] (-2+0.85,0) [partial ellipse = 0:50:0.85 and 0.25];
\draw[thick, dashed] (-2+0.85,0) [partial ellipse = 130:180:0.85 and 0.25];
\draw[thick] (0,0) [partial ellipse = 0:360:1.7 and 2.7];
\draw[thick] (0.25,0) [partial ellipse = 100:260:0.8 and 1.3];
\draw[thick] (-0.15,0) [partial ellipse = -83:83:0.7 and 1.2];
\draw[thick] (-1.125,0) [partial ellipse = 180:360:0.575 and 0.125];
\node[above] at (-1.125,0) {$d$};
\end{scope}
\draw[thick, -latex] (-1,0) -- (1,0);
\node[above] at (0,0) {$d\gg 1/\varepsilon$};
\begin{scope}[xshift = 4cm]
\draw[thick] (0,0) [partial ellipse = 0:360:2 and 3];
\draw[thick] (0.2,0) [partial ellipse = 100:260:0.5 and 1];
\draw[thick] (-0.2,0) [partial ellipse = -70:70:0.4 and 0.9];
\draw[thick, dashed, -latex] [partial ellipse = 130:-50:1.2 and 2.25];
\draw[thick, dashed, -latex] [partial ellipse = -50:-230:1.2 and 2.25];
\end{scope}
\end{tikzpicture}
\caption{A string in thermal $\text{AdS}_3$ which wraps the contractible cycle many times behaves like a massive particle.}
\label{fig:massive-particle}
\end{figure}
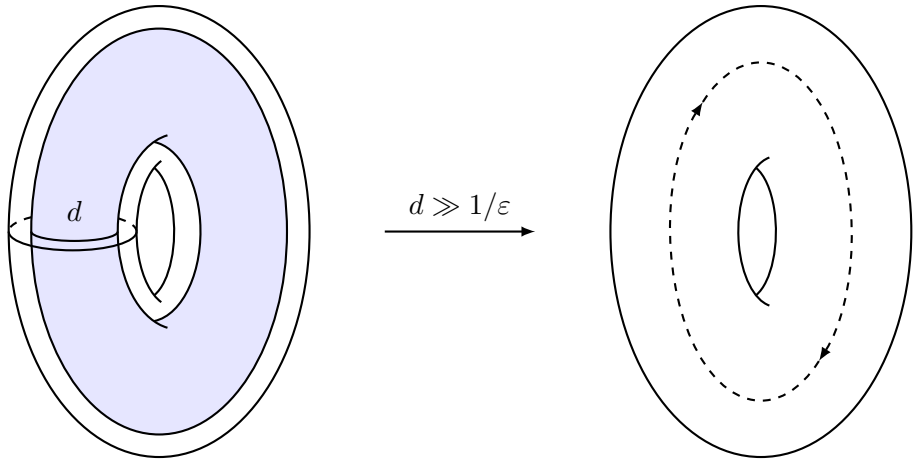

As a simple example, let us consider the classical solutions which arise when computing the one-loop partition function of the string theory. We consider $\gamma$ to take values on a torus, so that the target space geometry is thermal $\text{AdS}_3$ with modular parameter $t$. Specifically, that means we consider the identification $\gamma\sim\gamma+1$ and $\gamma\sim\gamma+t$. Furthermore, we take the worldsheet to be a torus with modular parameter $\tau$. As explained in Section \ref{sec:partition-function}, the appropriate covering map is now a holomorphic map from the worldsheet torus to the boundary torus of thermal $\text{AdS}_3$ and takes the form $\Gamma(z)=dz$ for some positive integer $d$. Furthermore, we demand $\tau=(at+b)/d$ for integers $a,b$ in order for this map to be well-defined, i.e. so that $\gamma(z)\sim\gamma(z+1)\sim\gamma(z+\tau)$ on the $\text{AdS}_3$ boundary torus. Here, $a$ represents the number of times the worldsheet wraps the compact time direction of the $\text{AdS}_3$ boundary, and $d$ counts the number of times the worldsheet wraps the angular direction. For this covering map, the radial profile is given by
\begin{equation}
r^2(z,\bar{z})=\varepsilon^2|\partial\Gamma|^2=\varepsilon^2d^2\,.
\end{equation}
If we consider the limit in which the worldsheet wraps the angular direction many times, i.e. $d\gg 1/\varepsilon$, we see that the radial profile vanishes, i.e. the string is completely localised to the center of $\text{AdS}_3$. In this limit, the string behaves like a massive particle in the center of $\text{AdS}_3$, which sources a conical defect in the resulting geometry \cite{Eberhardt:2021jvj}, see Figure \ref{fig:massive-particle}. Note that the existence of conical defects in an effective $\text{AdS}_3$ bulk theory is of interest from a purely gravitational stand-point (see, for instance, \cite{Arefeva:2014aoe,Berenstein:2022ico}), and it has been argued that the inclusion of these singularities is required for the consistency of the gravitational path integral \cite{Benjamin:2020mfz}. Thus, although our discussion is not related to pure $\text{AdS}_3$ gravity, it is satisfying to see the existence of conical defects arising automatically from a worldsheet theory.

We should emphasize that this limit is very strange from a physical perspective. The parameter $1/\varepsilon$ represents the UV cutoff in the dual CFT, and the duals of strings with winding $w$ are states whose conformal weights grow linearly in $w$. Thus, taking $w_i\gg 1/\varepsilon$ corresponds in the dual CFT to considering states whose energies are higher than the UV cutoff, which is physically not a meaningful thing to do. Furthermore, the solutions \eqref{eq:Wakimoto-classical} only solve the equations of motion in the limit $\varepsilon\to 0$, and we have no reason to expect $w_i\gg 1/\varepsilon$ to be a physically meaningful limiit from the classical sigma model. However, as we will see below, this limit is useful in explaining schematically how the semiclassical Nambu-Goto-like action \eqref{eq:Strebel-Liouville} emerges from a string moving in $\text{AdS}_3$. Thus, we will go forward with considering correlators in the limit $w_i\gg 1/\varepsilon$, while remembering that this limit is physically poorly defined, and that we should take it with a rather large grain of salt.

\subsection{The Strebel metric from classical geometry}

Consider the semiclassical solutions \eqref{eq:Wakimoto-classical} of the previous section, and consider Poincar\'e coordinates $(r,\gamma,\bar{\gamma})$ on $\text{AdS}_3$. As we mentioned above, the semiclassical solutions give us an embedding of the string into $\text{AdS}_3$ via
\begin{equation}
r(z,\bar{z})=e^{-\Phi(z,\bar{z})}\,,\quad \gamma(z,\bar{z})=\Gamma(z)\,,\quad \bar{\gamma}(z,\bar{z})=\bar{\Gamma}(\bar{z})\,.
\end{equation}
Now, the Poincar\'e metric $\mathrm{d}s^2=(\mathrm{d}r^2+\mathrm{d}\gamma\,\mathrm{d}\bar{\gamma})/r^2$ can be pulled back to the string worldsheet using this embedding. The result is
\begin{equation}
\mathrm{d}s_{\text{worldsheet}}^2=(\mathrm{d}\Phi)^2+e^{2\Phi}|\partial\Gamma|^2\mathrm{d}z\,\mathrm{d}\bar{z}\,.
\end{equation}
Using the relation $r^2=\varepsilon^2|\partial\Gamma|^2$ from \eqref{eq:Wakimoto-classical}, we end up with
\begin{equation}
\mathrm{d}s_{\text{worldsheet}}^2=\left(\frac{1}{4}\left|\frac{\partial^2\Gamma}{\partial\Gamma}\right|^2+\frac{1}{\varepsilon^2}\right)\mathrm{d}z\,\mathrm{d}\bar{z}\,.
\end{equation}
In the usual limit $\varepsilon\to 0$, the worldsheet metric is then conformally equivalent to the flat $\mathrm{d}z\,\mathrm{d}\bar{z}$ metric. However, the Strebel differential
\begin{equation}
\varphi=\frac{2}{N^2}S[\Gamma]\sim-\frac{1}{N^2}\left(\frac{\partial^2\Gamma}{\partial\Gamma}\right)^2
\end{equation}
is well-defined as $N\to\infty$, and thus we can write the worldsheet metric in this limit as
\begin{equation}
\mathrm{d}s^2_{\text{worldsheet}}\sim\left(\frac{N^2}{4}|\varphi|+\frac{1}{\varepsilon^2}\right)\mathrm{d}z\,\mathrm{d}\bar{z}\,.
\end{equation}
Therefore, if we take the double scaling limit $\varepsilon\to 0$ and $N\varepsilon\to\infty$, we see that the metric is dominated by the radial profile of the string motion, and we are left with the Strebel metric
\begin{equation}
\mathrm{d}s_{\text{worldsheet}}^2\sim N^2\mathrm{d}s^2_{\text{Strebel}}=\frac{N^2}{4}|\varphi|\mathrm{d}z\,\mathrm{d}\bar{z}\,.
\end{equation}
In this sense, the Strebel metric is the pullback of the $\text{AdS}_3$ metric precisely in the large-twist limit. Thus, we see that the expression \eqref{eq:Strebel-Liouville} for correlators in the symmetric orbifold can be recovered geometrically as the semiclassical motion of a string moving in $\text{AdS}_3$ in the large-twist limit, which was predicted in \cite{Gaberdiel:2020ycd}. In this limit, the Strebel metric comes entirely from the radial part of the $\text{AdS}_3$ metric, and so we see that the worldsheet moves almost purely in the radial direction of $\text{AdS}_3$.

In practice, one wants to compute the Nambu-Goto action
\begin{equation}
S_{\text{NG}}=\frac{1}{4\pi\alpha'}\int_{\Sigma}\frac{N^2}{4}|\varphi|\,,
\end{equation}
where we have briefly reintroduced the $\alpha'$ dependence. Since the area of the worldsheet diverges as $N\to\infty$, it is convenient to introduce an effective parameter $\alpha'_{\text{eff}}=\alpha'/N^2$ so that
\begin{equation}
S_{\text{NG}}=\frac{1}{4\pi\alpha'_{\text{eff}}}\int_{\Sigma}\frac{1}{4}|\varphi|\,,
\end{equation}
which describes a string with effective tension $T=N^2/\alpha'$. The `tensionless' limit of $\text{AdS}_3$ string theory is obtained by taking $\alpha'=1$ (in units of the $\text{AdS}_3$ radius). The effect of the large-twist limit is thus to introduce an effective large string tension, thus allowing the use of semiclassical methods to study the worldsheet.

\begin{figure}
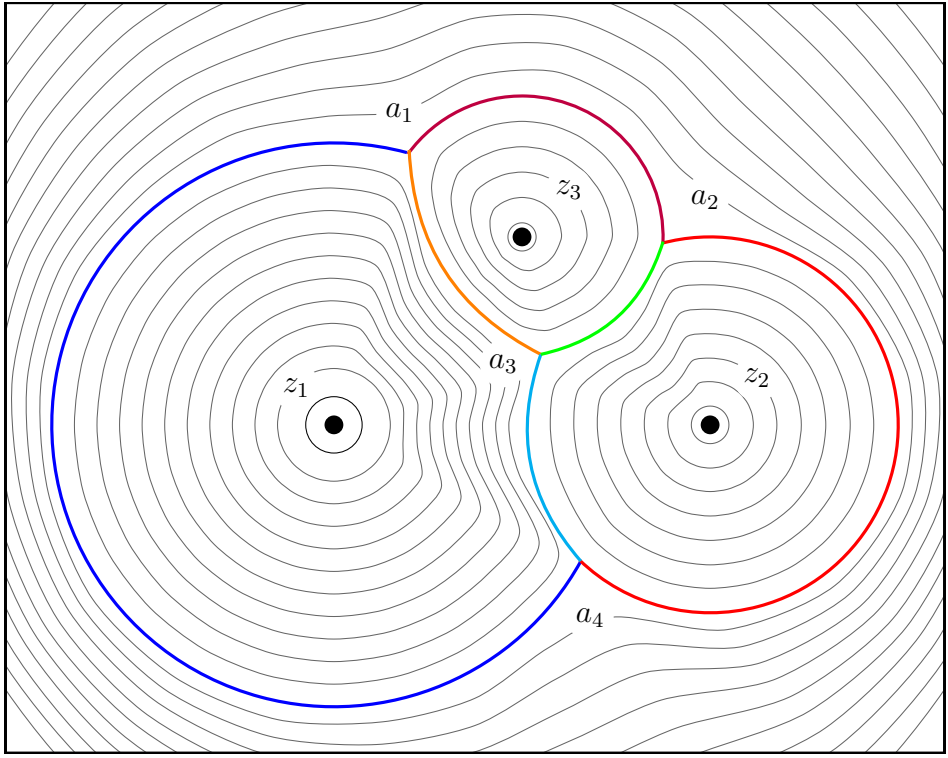

\centering

\caption{The natural coordinate system associated to a Strebel differential $\varphi$ on the worldsheet. Contours correspond to lines of constant time in the local $(y,\bar{y})$ coordinates near each insertion point. Colored lines correspond to the loci where the local coordinates have to be glued together. The zeroes $a_i$ of the Strebel differential lie on these so-called `critical' trajectories. Figure taken from \cite{Gaberdiel:2020ycd}.}
\label{fig:trajectories}
\end{figure}

Interestingly, there is a natural geometric interpretation of the Strebel gauge metric, which has wide application in string theory and string field theory \cite{Zwiebach:1990nh,Moeller:2004yy}. Given a Strebel differential $\varphi$, we can locally define `natural' worldsheet coordinates $\mathrm{d}y=\frac{N}{2}\sqrt{\varphi}$ and $\mathrm{d}\bar{y}=\frac{N}{2}\sqrt{\bar{\varphi}}$, which are defined on $\Sigma\setminus\{z_1,\ldots,z_n\}$.\footnote{The worldsheet coordinate $y$ is not to be confused with the spectral curve of Section \ref{sec:large-twist}.} In these coordinates, the worldsheet metric is given simply by
\begin{equation}
\mathrm{d}s^2_{\text{worldsheet}}=\mathrm{d}y\,\mathrm{d}\bar{y}\,,
\end{equation}
i.e. the natural coordinates $(y,\bar{y})$ locally describe a flat worldsheet metric. Near a pole of the Strebel differential located at the insertion point $z_i$ in the original coordinates, we have
\begin{equation}
\mathrm{d}y=\frac{N}{2}\sqrt{\varphi}\sim\frac{w_i/2}{(z-z_i)}\mathrm{d}z\,,
\end{equation}
so that
\begin{equation}
y\sim\frac{w_i}{2}\log(z-z_i)\,.
\end{equation}
Near $z=z_i$, the worldsheet looks like a semi-infinte tube of circumference $w_i/2$, and the $(y,\bar{y})$ coordinates are the natural cylindrical coordinates on this semi-infinite tube. The $(y,\bar{y})$ coordinates are only locally defined near the poles $z_i$ of $\varphi$, and away from the poles the different coordinate systems have to be `glued' together along the so-called `critical' trajectories of $\varphi$ (see Figure \ref{fig:trajectories}).

As was shown in \cite{Zwiebach:1990nh}, the Strebel-gauge metric can be seen to solve a minimal area problem. In particular, the worldsheet metric $\mathrm{d}s^2=N^2|\varphi|/4$ is the minimal-area metric among those satisfying the following three criteria:
\begin{itemize}

	\item The metric is induced by some quadratic differential $\varphi$.

	\item Given a puncture $z_i$ and a path $\gamma_i$ surrounding $z_i$, the length of $\gamma_i$ is fixed and given by $2\pi N\alpha_i/4=\pi(w_i-1)/2$.

	\item The metric has at most quadratic poles at $z=z_i$.

\end{itemize}
The second condition translates in the language of $\text{AdS}_3$ to the statement that the worldsheet winds $w_i$ times around its insertion point. The third allows one to define a regularised area of $\Sigma\setminus\{z_1,\ldots,z_n\}$ (so that the minimal area problem is well-defined).\footnote{The first condition, that the worldsheet metric is induced by a quadratic differential, does not, as far as the author knows, have a justification in semiclassical string theory. We thus take it as an assumption.} Thus, the Strebel gauge metric can be thought of as the minimal area of a string propagating in $\text{AdS}_3$ which winds around its insertion points $w_i$ times. In this way, we can really think of the Strebel-gauge metric as the semiclassical worldsheet minimizing the Nambu-Goto action, thus justifying the stringy interpretation of \eqref{eq:Strebel-Liouville}.

In the $(y,\bar{y})$ coordinate system, we can calculate the semiclassical motion of the string in the $(r,\gamma,\bar{\gamma})$ $\text{AdS}_3$ coordinates rather easily. In terms of the covering map, we have
\begin{equation}
\mathrm{d}y\sim i\frac{1}{2}\partial\log{\partial\Gamma}
\end{equation}
and so $\partial\Gamma\sim e^{-2iy}$. Since $r^2=\varepsilon^2|\partial\Gamma|^2$, we have
\begin{equation}
r^2(y,\bar{y})=\varepsilon^2e^{4\text{Im}(y)}\,.
\end{equation}
Thus, as $\text{Im}(y)\to-\infty$, we see that $r^2\to 0$ and the worldsheet approaches the boundary of $\text{AdS}_3$. As we increase $y$, $r$ increases and the worldsheet approaches the center of $\text{AdS}_3$. Thus, near $z=z_i$, we see again that the worldsheet resembles a semi-infinite tube which extends from the boundary of $\text{AdS}_3$ deep into the bulk (see Figure \ref{fig:tubes}). Each of these tubes corresponds to a face in the graph of Figure \ref{fig:trajectories}, and they are glued together in the $\text{AdS}_3$ bulk along the graph of critical trajectories of $\varphi$.

It is worth noting that the above coordinate system has a natural place in the context of string field theory, where it is used as a preferred coordinate for the insertion of off-shell states on the worldsheet \cite{Moeller:2004yy}. It would be interesting, then, to explore whether the large-twist limit of the tensionless worldsheet theory could be used to explore closed-string correlators of off-shell states. This could potentially provide a playground for understanding closed-string field theory for the tensionless string.

\section{Relation to the worldsheet theory}\label{sec:reconstruction}

\subsection{The free field realisation}

The worldsheet theory \cite{Eberhardt:2018} which is dual to the symmetric orbifold $\text{Sym}^K(\mathbb{T}^4)$ is most naturally written in the hybrid formalism of Berkovits, Vafa, and Witten \cite{Berkovits_1999}, in which it is expressed in terms of a sigma model on the supergroup $\text{PSU}(1,1|2)_{L}\times\text{PSU}(1,1|2)_{R}$ times a (topologically twisted) sigma model on $\mathbb{T}^4$. The resulting WZW model has current algebra $\mathfrak{psu}(1,1|2)_k\oplus\mathfrak{psu}(1,1|2)_k$, and the tensionless limit is achieved by taking $k=1$. One reason for the tractibility of the $k=1$ theory is that the superalgebra $\mathfrak{psu}(1,1|2)_1$ enjoys a realisation in terms of free fields \cite{Eberhardt:2018,Dei:2020}. Let us focus on the bosonic subalgebra $\mathfrak{sl}(2,\mathbb{R})_1\subset\mathfrak{psu}(1,1|2)_1$, since it contains the salient features for our discussion. The free field realisation consists of two spin $-\frac{1}{2}$ bosons $(\lambda,\mu)$ and two canonically conjugate spin $\frac{3}{2}$ bosons $(\mu^{\dagger},\lambda^{\dagger})$ which obey the OPEs
\begin{equation}
\lambda(z)\,\mu^{\dagger}(w)\sim\frac{1}{z-w}\,,\quad\mu(z)\,\lambda^{\dagger}(w)\sim\frac{1}{z-w}\,.
\end{equation}
This theory has two $\mathfrak{u}(1)$ symmetries obtained by simultaneous rescaling of the fields. In particular, one such symmetry is given by
\begin{equation}\label{eq:scaling-symmetry}
\begin{split}
(\lambda,\mu)&\to\alpha(\lambda,\mu)\,,\\
(\mu^{\dagger},\lambda^{\dagger})&\to\alpha^{-1}(\mu^{\dagger},\lambda^{\dagger})\,,
\end{split}
\end{equation}
and is generated by a current $U$. In the full worldsheet theory, one gauges this global symmetry, and the gauge invariant bilinears of the free fields generate the current algebra $\mathfrak{sl}(2,\mathbb{R})_1$. We review this free field theory in detail in Appendix \ref{sec:free-fields}.

Geometrically, due to the symmetry \eqref{eq:scaling-symmetry}, we can think of the resulting description as a (first order) sigma model on $(\lambda,\mu)/\sim\,\,=\mathbb{CP}^1$. The daggered fields $(\mu^{\dagger},\lambda^{\dagger})$ can then be thought of as the conjugate momenta of the $\mathbb{CP}^1$ coordinates $(\lambda,\mu)$. Conceptually, we think of $(\lambda,\mu)$ as parametrising the twistor space of two-dimensional field theory, and thus we will refer to them as twistors. In the full free field construction of $\mathfrak{psu}(1,1|2)_1$, we also include Fermions $\psi^a$ of spin $-\frac{1}{2}$ and their conjugates $(\psi^{\dagger})_a$ of spin $\frac{3}{2}$ and gauge the symmetry
\begin{equation}
\begin{split}
(\lambda,\mu)&\to\alpha(\lambda,\mu)\,,\\
(\mu^{\dagger},\lambda^{\dagger})&\to\alpha^{-1}(\mu^{\dagger},\lambda^{\dagger})\,,\\
\psi^{a}&\to \alpha\psi^a\,,\\
\quad(\psi^{\dagger})_a&\to\alpha^{-1}(\psi^{\dagger})_a\,.
\end{split}
\end{equation}
The full supersymmetric model then has coordinates $(\lambda,\mu|\psi^{a})$ which parametrise the super-twistor space $\mathbb{CP}^{1|2}$.

The relationship between the `twistor' worldsheet fields and the usual description of $\text{AdS}_3$ in terms of a WZW model on $\text{SL}(2,\mathbb{R})$ is that one can generate the $\text{SL}(2,\mathbb{R})$ currents in terms of bilinears of the twistor fields with conformal weight $h=1$. Specifically, we can identify
\begin{equation}
J^+=\lambda^{\dagger}\lambda\,,\quad J^-=\mu^{\dagger}\mu\,,\quad J^3=\frac{1}{2}(\lambda^{\dagger}\mu-\mu^{\dagger}\lambda)\,.
\end{equation}
There is one more linearly independent combination that can be constructed, and it generates the gauge symmetry \eqref{eq:scaling-symmetry}, see Appendix \ref{sec:free-fields}.

A remarkable feature of the worldsheet theory is that its correlation functions localise to configurations for which there is a holomorphic covering $\Gamma:\Sigma\to\mathbb{CP}^1$. In fact, the free fields $(\lambda,\mu)$ provide an explicit representation of this map. Given vertex operators $V^{w_i}(x_i,z_i)$ describing the emission of a string with worldsheet coordinate $z_i$ and $\partial\text{AdS}_3$ coordinate $x_i$, we have (see Appendix \ref{sec:free-fields})
\begin{equation}\label{eq:twistor-classical-correlators}
\begin{split}
\Braket{\lambda(z)\prod_{i=1}^{n}V^{w_i}(x_i,z_i)}&=\frac{1}{\sqrt{\partial\Gamma(z)}}\\
\Braket{\mu(z)\prod_{i=1}^{n}V^{w_i}(x_i,z_i)}&=\frac{\Gamma(z)}{\sqrt{\partial\Gamma(z)}}\,.
\end{split}
\end{equation}
The interpretation of this statement is that $\lambda=1/\sqrt{\partial\Gamma}$ and $\mu=\Gamma/\sqrt{\partial\Gamma}$ are the `classical' worldsheet coordinates in the $\text{AdS}_3$ boundary. Furthermore, unless the moduli of $\Sigma\setminus\{z_1,\ldots,z_n\}$ are situated in such a way that the covering map $\Gamma$ exists, the correlators of $V^{w_i}$ vanish and equation \eqref{eq:twistor-classical-correlators} is vacuous. When the covering map does exist, the pair $[\lambda:\mu]=[1:\Gamma]/\sqrt{\partial\Gamma}$ is then a choice of (non-canonical) presentation of this map on $\mathbb{CP}^1$. It can be thought of as the pullback of the `gauge-fixed' coordinate $[1:x]$ onto the worldsheet by $\Gamma$, i.e.
\begin{equation}
[\lambda:\mu]=\Gamma^*\left(\frac{[1:x]}{\sqrt{\mathrm{d}x}}\right)\,,
\end{equation}
where $\Gamma^*$ denotes the pullback from the appropriate holomorphic line bundle of $\mathbb{CP}^1$ to the corresponding line bundle on $\Sigma$.

\subsection{Twistors from the Strebel differential}

As it turns out, the worldsheet free fields are deeply connected to the Strebel differential. Given a covering map $\Gamma$, we can consider the holomorphic differential equation
\begin{equation}\label{eq:schrodinger-equation}
\partial^2f+\frac{1}{2}S[\Gamma]f=0\,.
\end{equation}
This equation has a two-dimensional family of solutions, which, under coordinate transformations, transform as conformal primaries of weight $h=-\frac{1}{2}$ (i.e. define sections of the inverse spinor bundle $S^{-1}$ for some given spin structure $S$). Indeed, one can immediately write down a pair of solutions for $f$, which are precisely the classical values of the twistor fields, namely
\begin{equation}\label{eq:free-fields-classical}
\lambda=\frac{1}{\sqrt{\partial\Gamma}}\,,\quad\mu=\frac{\Gamma}{\sqrt{\partial\Gamma}}\,.
\end{equation}
The fact that the twistor fields \eqref{eq:free-fields-classical} satisfy the differential equation \eqref{eq:schrodinger-equation} can be seen either from directly verifying that \eqref{eq:free-fields-classical} is a solution, or by factorising \eqref{eq:schrodinger-equation} as
\begin{equation}
\left(\partial^2+\frac{1}{2}S[\Gamma]\right)f=\left(\partial-\partial\log\sqrt{\partial\Gamma}\right)\left(\partial+\partial\log\sqrt{\partial\Gamma}\right)f\,,
\end{equation}
so that $\lambda$ is the solution obtained by demanding $f$ is annihilated by $\partial+\partial\log\sqrt{\partial\Gamma}$, and $\mu$ is obtained similarly.

Recalling that, in the large-twist limit, the Strebel differential is related to the Schwarzian via
\begin{equation}
\varphi=-\frac{2}{N^2}S[\Gamma]\,,
\end{equation}
we see that the fields $\lambda$ and $\mu$ can be constructed through the solutions of the differential equation
\begin{equation}\label{eq:strebel-schrodinger}
\partial^2f=\frac{N^2}{4}\varphi\,f\,.
\end{equation}
Thus, there is a natural relationship between the worldsheet free fields and the Strebel differential, namely
\begin{equation}
\frac{\partial^2\lambda}{\lambda}=\frac{\partial^2\mu}{\mu}=\frac{N^2}{4}\varphi\,.
\end{equation}

If we know the explicit form of the covering map $\Gamma$, this differential equation is easy to solve. However, in the large-twist limit, $\Gamma$ itself is not a well-defined function, and the best option we have is to implicitly define it through the Strebel differential. We can similarly recover the worldsheet fields $\mu,\lambda$ in this limit purely through the Strebel differential. To see this, note that the holomorphic Schr\"odinger equation \eqref{eq:strebel-schrodinger} can be solved via the WKB approximation in the limit $N\gg 1$. Indeed, in a neighbourhood of a regular point $z_0$ (i.e. $z_0$ is not a critical point of $\Gamma$), one can immediately write down a pair of approximate solutions given by
\begin{equation}\label{eq:WKB-approximation}
f_{\pm}=\varphi^{-1/4}\exp\left(\pm \frac{N}{2}\int_{z_0}^{z}\sqrt{\varphi}\right)\,.
\end{equation}
This solution, however, is not valid as we take $z$ very far away from $z_0$. In particular, if $z_0$ is within one of the regions bounded by the critical trajectories of $\varphi$ (see Figure \ref{fig:trajectories}), then the above approximation will only be valid within that region. Once $z$ crosses one of the critical trajectories, $\sqrt{\varphi}$ crosses a zero or a branch cut, and the WKB approximation breaks down (analogous to when one encounters a `turning point' in one-dimensional quantum mechanics). That is, the solution to the WKB approximation is only valid locally near $z=z_i$. A full solution would require gluing the separate solutions together.

Within each region surrounding the insertion, we can interpret the two solutions above as the combinations $\lambda(z)$ and $\mu(z)-x_i\lambda(z)$. Indeed, near $z=z_i$ we have
\begin{equation}
f_{\pm}\sim(z-z_i)^{1/2}\exp\left(\pm\frac{w_i}{2}\int_{z_0}^{z}\frac{1}{z-z_i}\mathrm{d}z\right)\sim(z-z_i)^{\frac{1}{2}\pm\frac{w_i}{2}}\,,
\end{equation}
which is precisely the asymptotic behaviour of $\mu(z)-x_i\lambda(z)$ and $\lambda(z)$, respectively.

We note that we can write the above solutions in terms of the natural coordinate system $(y,\bar{y})$ induced by $\varphi$. Given a base-point $z_0$, we have
\begin{equation}
y(z)=\frac{N}{2}\int_{z_0}^{z}\sqrt{\varphi}\,,\qquad\bar{y}(\bar{z})=\frac{N}{2}\int_{\bar{z}_0}^{\bar{z}}\sqrt{\bar{\varphi}}\,.
\end{equation}
In this coordinate system, the metric is simply $\mathrm{d}s^2=\mathrm{d}y\,\mathrm{d}\bar{y}$, and is flat unless $\varphi$ is singular. The solutions $f_{\pm}$ then take the very simple form
\begin{equation*}
f_{\pm}(y)\sim \frac{e^{\pm y}}{\sqrt{\mathrm{d}y}}\,.
\end{equation*}
Since we think of $f_{\pm}$ as giving local coordinates on the target space $\mathbb{CP}^1$, we can take their ratio to get a local coordinate on $\mathbb{C}\cup\{\infty\}$, namely
\begin{equation}
\frac{f_+}{f_-}=e^{2y}\sim(z-z_i)^{w_i}\,.
\end{equation}
Of course, this ratio is nothing other that $\Gamma(z)-x_i$, since $f_+=\mu-x_i\lambda$ and $f_-=\lambda$. The WKB approximation then simply tells us that the coordinates $y$ and the coordinates $\Gamma-x_i$ are related to each other by exponentiation.

Finally, let us comment that there is a suggestive relationship between the twistor fields and the Liouville field $\Phi$. Recall that the relationship between the Strebel differential $\varphi$ and $\Phi$ is given by $\partial\Phi=iN\sqrt{\varphi}$, so that
\begin{equation}
\Phi=iN\int_{z_0}^{z}\sqrt{\varphi}\,.
\end{equation}
This allows us to instantly write down the solutions $f_{\pm}$ in terms of the Liouville field, namely
\begin{equation}
f_{\pm}\sim (\partial\Phi)^{-1/2}e^{\pm\Phi/2}\,.
\end{equation}
Since, within the coordinate system near $z=z_i$, the solutions $f_{\pm}$ are related to the free fields $\lambda,\mu$, we can think of $\Phi$ as the bosonisation of these free fields.\footnote{Similar expressions relating the $\text{AdS}_3$ free field to the Liouville field $\Phi$ are found in, for example, equation (2.9) of \cite{Bhat:2021dez}. The difference between their expression and ours is a choice of gauge in the free field realisation.}

Let us set $x_i=0$ for convenience, so that we can identify the two solutions $f_{\pm}$ with $\mu$ and $\lambda$. Then we have explicitly
\begin{equation}
\lambda\sim\frac{e^{-\Phi/2}}{\sqrt{\partial\Phi}}\,,\quad\mu\sim\frac{e^{\Phi/2}}{\sqrt{\partial\Phi}}\,.
\end{equation}
Now, we can propose a similar relationship between the Liouville field $\Phi$ and the daggered fields $\mu^{\dagger},\lambda^{\dagger}$. In particular, it was noted in \cite{Eberhardt:2019,Bhat:2021dez} that $\Phi$ can be thought of as the bosonisation of $J^3$ in the $\mathfrak{sl}(2,\mathbb{R})$ current algebra, i.e. $J^3=\partial\Phi$, and so naively we can propose
\begin{equation}
\lambda^{\dagger}\mu-\mu^{\dagger}\lambda=2J^3=2\partial\Phi\,.
\end{equation}
Furthermore, since we are imposing the gauge symmetry \eqref{eq:scaling-symmetry}, we further propose that the current corresponding to this symmetry vanishes \cite{Bhat:2021dez}. Specifically,
\begin{equation}
\lambda^{\dagger}\mu+\mu^{\dagger}\lambda=0\,.
\end{equation}
Thus, we can immediately write down the semiclassical values of $\mu^{\dagger},\lambda^{\dagger}$ to be
\begin{equation}
\mu^{\dagger}=-(\partial\Phi)^{3/2}e^{\Phi/2}\,,\quad\lambda^{\dagger}=(\partial\Phi)^{3/2}e^{-\Phi/2}\,.
\end{equation}
Furthermore, we can calculate the semiclassical stress-tensor of these expressions, and we find
\begin{equation}
T=\frac{1}{2}\left(\mu^{\dagger}\partial\lambda-\lambda\partial\mu^{\dagger}+\lambda^{\dagger}\partial\mu-\mu^{\dagger}\partial\lambda\right)+\cdots\sim(\partial\Phi)^2\,,
\end{equation}
where $\cdots$ represents terms which vanish under the gauge constraint. Since $(\partial\Phi)^2$ is simply the Strebel differential (up to a constant), we reproduce the semiclassical result of \cite{Bhat:2021dez} that the stress tensor of the free field theory is related to the Strebel differential.\footnote{As was pointed out in \cite{Bhat:2021dez}, this is not so surprising, and in general stress-tensors of Liouville theories and Strebel differentials converge in the `large-twist' limit, c.f. the discussion in Chapter 5 of \cite{eynard2018lectures}.}

\subsection{Reconstructing the worldsheet}

Above, we argued for a suggestive relationship between the Strebel differential appearing in the large-twist limit of the symmetric orbifold CFT and the twistorial free field description of the worldsheet. However, let us for a moment assume we know nothing of the worldsheet theory. Given nothing more than the knowledge that correlators are naturally expressed in terms of Strebel differentials, how much about the worldsheet theory can we extract? 

Let $\varphi$ be a generic Strebel differential on the surface $\Sigma$ with quadratic residues $\alpha_i^2$ at $z=z_i$. Armed with the hindsight of the above discussions, we might simply postulate that the worldsheet theory can be described in terms of the solutions to the holomorphic Schr\"odinger equation
\begin{equation}
\partial^2f=\varphi\,f\,.
\end{equation}
What can we say about the solutions $f$? Near a pole $z=z_i$ we have
\begin{equation}
\partial^2f\sim\frac{\alpha_i^2f}{(z-z_i)^{2}}\,.
\end{equation}
The independent solutions to this differential equation are
\begin{equation}\label{eq:sol-asymptotics}
f_{\pm}(z)\sim (z-z_i)^{\frac{1\pm r_i}{2}}\,,\quad r=\sqrt{1+\alpha_i^2}\,.
\end{equation}
Thus, since $\alpha_i^2\in\mathbb{R}_+$ by construction, one of the solutions has a zero of degree $\frac{1+r_i}{2}$ at $z=z_i$ and the other has a pole of degree $\frac{r_i-1}{2}$. If $f_{\pm}$ are truly to describe worldsheet degrees of freedom, then the asymptotic behaviours \eqref{eq:sol-asymptotics} should describe their OPEs with vertex operators $V_{\alpha_i}(z_i)$ on the worldsheet. In particular, promoting $f_{\pm}$ to worldsheet fields, we should have
\begin{equation}\label{eq:deduced-OPE}
f_{\pm}(z)\,V_{\alpha_i}(z_i)\sim (z-z_i)^{\frac{1\pm r_i}{2}}\,.
\end{equation}
If we assume the existence of a current $\partial\phi$ on the worldsheet under which $f_{\pm}$ have charges $\pm\frac{1}{2}$, then we can explicitly construct such a vertex operator as\footnote{For the tensionless $\text{AdS}_3$ theory, the appropriate current is $\partial\phi=J^3$, and the exponential $e^{r_i\phi/2}$ is the spectral flow operator, see Appendix \ref{sec:free-fields}.}
\begin{equation}
V_{\alpha_i}(z)\sim e^{r_i\phi/2}\,,
\end{equation}
The remaining piece of the OPE then tells us that $f_{\pm}(z)\ket{\Omega}\sim z^{\frac{1}{2}}$ as $z\to 0$, where $\ket{\Omega}$ is the CFT vacuum. This is the hallmark of a conformal field of weight $h=-\frac{1}{2}$. The simplest way to build a worldsheet CFT is then to introduce canonical conjugate fields for the two $h=-\frac{1}{2}$ fields, which are of weight $h=\frac{3}{2}$.

Thus, from the (as yet unjustified, but motivated by the example of the tensionless string) postulate that the worldsheet degrees of freedom are given by solutions to the holomorphic Schr\"odinger equation with $\varphi$ as the potential, we can deduce the OPEs \eqref{eq:deduced-OPE} and the fact that the worldsheet fields have conformal weight $h=-\frac{1}{2}$, assuming that the vertex operators can be expressed as exponentials of a scalar $\phi$, under whose current $\partial\phi$ the worldsheet fields $f_{\pm}$ come with charge $\pm\frac{1}{2}$.

Of course, these properties are nearly those of the worldsheet dual of the symmetric orbifold CFT.\footnote{It is not obvious, however, how to see the gauging of the scaling symmetry in equation \eqref{eq:scaling-symmetry}.} However, it has been argued \cite{Gopakumar:2003ns,Gopakumar:2004qb,Gopakumar:2005fx} that the worldsheet dual to generic free (holographic) gauge theories should also be expressible naturally in terms of Strebel differentials. Thus, it would not be unreasonable to assume that the above discussion generalises to the worldsheet dual of \textit{any} holographic free CFT, not just the symmetric orbifold. The conclusion would be that the worldsheet dual of a generic free CFT is, at least partially, describable in terms of worldsheet fields which formally resemble those of the free field realisation of $\mathfrak{psu}(1,1|2)_1$. This is indeed true for the case of the worldsheet dual of free $\mathcal{N}=4$ super Yang-Mills \cite{Gaberdiel:2021qbb,Gaberdiel:2021jrv}, and has also recently shown to be true for a large class of free $\mathcal{N}=2$ quiver gauge theories dual to tensionless string theory on $\text{AdS}_5\times(\text{S}^5)/\mathbb{Z}_N$ \cite{Gaberdiel:2022iot}. It would thus be interesting to explore further the connection between twistor-like worldsheet theories and the stringy duals to free holographic CFTs.

\section{\boldmath Comments on \texorpdfstring{AdS$_2$}{AdS2}}\label{sec:ads2}

In this section, we explore the relationship between the tensionless string on $\text{AdS}_3$ in the large-twist limit and gravity on Euclidean $\text{AdS}_2$ via a particular dimensional reduction. In particular, we show that the dynamics of a string which ends on a certain type of D-brane in $\text{AdS}_3$ is seemingly governed by a one-dimensional Schwarzian action in the large-twist limit, suggesting a potential connection to Jackiw-Teitelboim gravity \cite{Jackiw:1984je} in two dimensions.

\subsection[Branes in \texorpdfstring{AdS$_3$}{AdS3}]{\boldmath Branes in \texorpdfstring{AdS$_3$}{AdS3}}

\begin{figure}
\centering
\begin{tikzpicture}
\begin{scope}[xshift = -4.5cm, scale = 0.75]
\draw[thick] (0,0) [partial ellipse = 167:373:4 and 1.33];
\draw[thick, dashed] (0,0) [partial ellipse = 0:180:4 and 1.33];
\fill[black, opacity = 0.1] (0,4) [partial ellipse = 0:360:3 and 1];
\draw[thick] (0,4) [partial ellipse = 180:360:3 and 1];
\draw[thick, dashed] (0,4) [partial ellipse = 0:180:3 and 1];
\draw[thick] (0,8) ellipse (4 and 1.33);
\draw[thick] (-3.904,0.29) to[out = 60, in = -90] (-3,4) to[out = 90, in = -60] (-3.904,7.71);
\draw[thick] (3.904,0.29) to[out = 120, in = -90] (3,4) to[out = 90, in = -120] (3.904,7.71);
\end{scope}
\begin{scope}[xshift = 4.5cm, scale = 0.75]
\draw[thick] (0,0) [partial ellipse = 167:373:4 and 1.33];
\draw[thick, dashed] (0,0) [partial ellipse = 0:180:4 and 1.33];
\fill[black, opacity = 0.1] (0,4) [partial ellipse = 0:360:3 and 1];
\draw[thick] (0,4) [partial ellipse = 180:360:3 and 1];
\draw[thick, dashed] (0,4) [partial ellipse = 0:180:3 and 1];
\fill[blue, opacity = 0.15] (-2.8,4) to[out = 90, in = -45] (-3.18,6) to[out = 45, in = -90] (-3.8,8) -- (3.8,8) to[out = -90, in = 135] (3.53,7) to[out = -135, in = 135] (3.03,5) to[out = -135, in = 90] (2.8,4) -- (-2.8,4);
\fill[blue, opacity = 0.15] (0,8) [partial ellipse = 180:0:3.8 and 1.2];
\fill[blue, opacity = 0.15] (0,4) [partial ellipse = 0:-180:2.8 and 0.8];
\draw[thick] (-2.8,4) to[out = 90, in = -45] (-3.18,6) to[out = 45, in = -90] (-3.8,8);
\draw[thick] (3.8,8) to[out = -90, in = 135] (3.53,7) to[out = -135, in = 135] (3.03,5) to[out = -135, in = 90] (2.8,4);
\draw[thick] (0,8) [partial ellipse = 360:0:3.8 and 1.2];
\draw[thick] (0,4) [partial ellipse = 0:-180:2.8 and 0.8];
\draw[thick] (0,8) ellipse (4 and 1.33);
\draw[thick] (-3.904,0.29) to[out = 60, in = -90] (-3,4) to[out = 90, in = -60] (-3.904,7.71);
\draw[thick] (3.904,0.29) to[out = 120, in = -90] (3,4) to[out = 90, in = -120] (3.904,7.71);
\node[left] at (-3.2,6) {$x_1$};
\node[right] at (3.53,7) {$x_2$};
\node[right] at (3.03,5) {$x_3$};
\end{scope}
\end{tikzpicture}
\caption{Left: A `spherical' brane localised at time $t=t_0$ in $\text{AdS}_3$. The induced metric on the brane is that of Euclidean $\text{AdS}_2$. Right: The worldsheet configuration associated to the CFT three-point function $\langle\mathcal{O}^{(w_1)}_1(x_1)\mathcal{O}^{(w_2)}_2(x_2)\mathcal{O}^{(w_2)}_2(x_2)\bket{\psi}$.}
\label{fig:brane}
\end{figure}
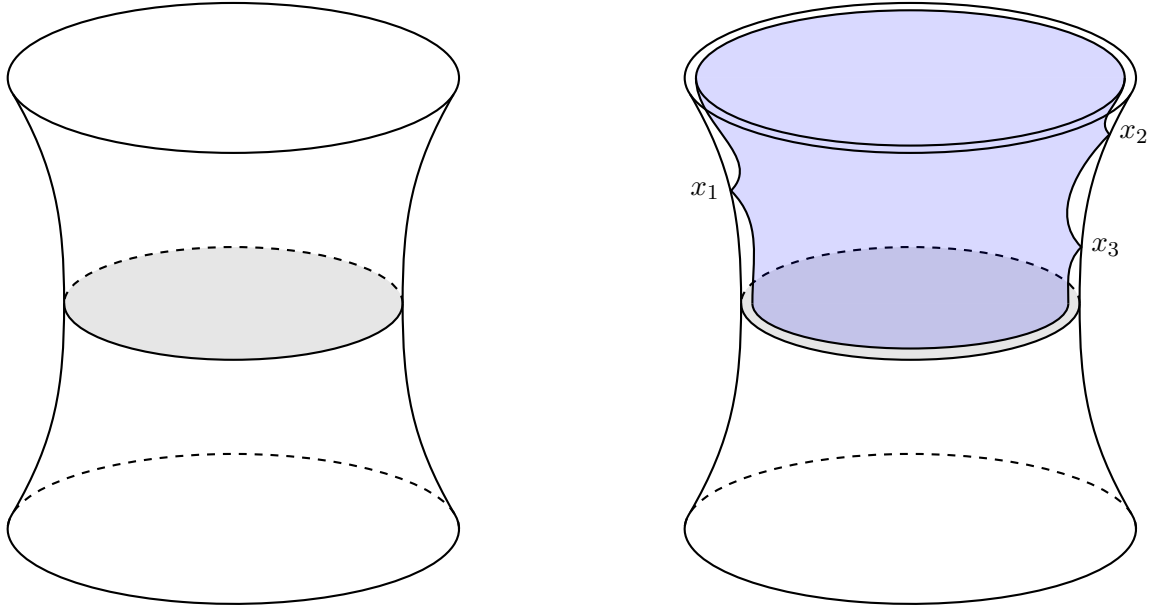

In \cite{Gaberdiel:2021kkp}, the duality between tensionless strings on $\text{AdS}_3\times\text{S}^3\times\mathbb{T}^4$ and the symmetric orbifold theory was extended to include gravitational backgrounds with D-branes. In particular, branes in the $\text{AdS}_3$ bulk which are localised at a specific point in time (so-called \textit{spherical branes} in the parlance of \cite{Bachas:2000fr}), depicted in Figure \ref{fig:brane}, were shown to be dual to boundary states in the symmetric orbifold theory which are `maximally symmetric', in the sense that they share the same boundary conditions in all copies of the $\mathbb{T}^4$ seed theory.

Let us denote such a boundary state in the symmetric orbifold as $\bket{\psi}$. We can consider correlation functions in the presence of such a boundary state
\begin{equation}\label{eq:boundary-correlator}
\Braket{\prod_{i=1}^{n}\mathcal{O}^{(w_i)}_i(x_i)}_{\psi,\mathbb{D}}:=\Bigg\langle\prod_{i=1}^{n}\mathcal{O}_{i}^{(w_i)}(x_i)\Bigg|
\hspace{-0.05cm}\Bigg|\psi\Bigg\rangle\hspace{-0.225cm}\Bigg\rangle_{\mathbb{D}}\,.
\end{equation}
Specifically, we model the base space of the CFT as the disk $\mathbb{D}=\{z\in\mathbb{C}:|z|\leq 1\}$ and impose the boundary condition corresponding to $\bket{\psi}$ on the boundary $|z|=1$. It was argued in \cite{Gaberdiel:2021kkp} that these correlation functions enjoy a similar construction in terms of covering maps $\Gamma:\Sigma\to\mathbb{D}$ which cover the disk $\mathbb{D}$ with some Riemann surface $\Sigma$ with boundary. Schematically, we can associate to each of these covering surfaces an Euler characteristic $\chi(\Sigma)=2-2g-n-b$, where $b$ is the number of boundary components of $\Sigma$, and the correlation function \eqref{eq:boundary-correlator} can be expressed as
\begin{equation}
\Braket{\prod_{i=1}^{n}\mathcal{O}^{(w_i)}_i(x_i)}_{\psi,\mathbb{D}}=\sum_{\Gamma:\Sigma\to\mathbb{D}}K^{\chi(\Sigma)}e^{-S_L[\Phi_{\Gamma}]}\Braket{\prod_{i=1}^{n}\mathcal{O}_i(z_i)}_{\psi,\Sigma}\,,
\end{equation}
where in the right-hand-side $z_i$ are the branch points of $\Gamma$ with $\Gamma(z_i)=x_i$, $\mathcal{O}_i$ are the seed theory operators defined on the covering surface $\Sigma$, and $S_{L}[\Phi_{\Gamma}]$ is the same Liouville action we introduced previously, except now defined on a Riemann surface with boundary
\begin{equation}
S_{L}[\Phi_{\Gamma}]=\frac{1}{8\pi}\int_{\Sigma}(2\partial\Phi_{\Gamma}\bar{\partial}\Phi_{\Gamma}+R\Phi)\,,
\end{equation}
and $\Phi_{\Gamma}$, as usual, is given by $\Phi_{\Gamma}=\log|\partial\Gamma|^2$.

Just as in the case of the symmetric orbifold on the sphere, there is a natural construction to associate a Strebel differential $\varphi$ to each covering space $\Gamma:\Sigma\to\mathbb{D}$. The simplest such method is to consider the compact double cover $\Sigma_c$ of $\Sigma$, for which there exists an orientation preserving diffeomorphism $\iota:\Sigma_c\to\Sigma_c$ such that $\Sigma\cong\Sigma_c/\iota$, where the boundary of $\Sigma$ is identified with the fixed-point set of $\iota$. We can then consider the covering map $\Gamma_c:\Sigma_c\to\mathbb{CP}^1$ which maps $z_i$ to $x_i$ and $\iota(z_i)$ to $-1/\bar{x}_i$ with branching $w_i$. Taking the degree of $\Gamma_c$ to be large, the Schwarzian derivative $S[\Gamma]$ converges to a Strebel differential $\varphi_c$ defined on $\Sigma_c$. Letting $i:\Sigma_c\to\Sigma$ be the projection map associated to $\iota$, we can then define a Strebel differential on $\Sigma$ by $i^*\varphi_c$.

By the arguments in Section \ref{sec:large-twist}, we can then relate the symmetric orbifold correlator to an integral over the moduli space $\mathcal{M}_{n,g,b}$ of surfaces of genus $g$ with $n$ punctures and $b$ boundary components\footnote{This is a moduli space of real dimension $\text{dim}_{\mathbb{R}}(\mathcal{M}_{g,n,b})=2n+6g-6+3b$.} as
\begin{equation}\label{eq:boundary-nambu-goto}
\Braket{\prod_{i=1}^{n}\mathcal{O}_i^{(w_i)}(x_i)}_{\psi,\mathbb{D}}\sim\sum_{g,b}K^{2-2g-n-b}\int_{\mathcal{M}_{n,g,b}}e^{-\frac{N^2}{4\pi}\int_{\Sigma}|\varphi|}\Braket{\prod_{i=1}^{n}\mathcal{O}_{i}(z_i)}_{\Sigma}\,.
\end{equation}
This has the holographic interpretation of an open string propagating in the background of Figure \ref{fig:brane}, where $N^2|\varphi|/4$ is the pullback of the effective $\text{AdS}_3$ metric.

\subsection{Dimensional reduction and the Schwarzian}

Let us now consider a particular dimensional reduction from $\text{AdS}_3$ to Euclidean $\text{AdS}_2$ in the above setup. We do this by treating the endpoints of the string as individual objects which move only in the plane $t=t_0$ and writing down their dynamics. The effect of the `dimensional reduction' on the covering map is to consider only the boundary covering map $\Gamma:\partial\Sigma\to\text{S}^1$ which maps the boundary of the worldsheet to the boundary circle of the spherical brane, and as such parametrizes the motion of the string endpoint. For now let us consider $b=1$ and $g=0$ (i.e. take the worldsheet topology to be a disk) since this is the leading contribution in $1/N$. Since the boundary circle can be thought of as the complex numbers with unit modulus, we have $\Gamma(\tau)=e^{i\theta(\tau)}$, where $\tau$ is the boundary angle on the covering space and $\theta$ is the angle on the Euclidean $\text{AdS}_2$ boundary circle.

In this limit, we can explicitly calculate the Strebel differential on the boundary circle. We have
\begin{equation}
\partial\Phi_{\Gamma}\,\bar{\partial}\Phi_{\Gamma}\to\left(\frac{\mathrm{d}}{\mathrm{d}\tau}\log\frac{\mathrm{d}\theta}{\mathrm{d}\tau}\right)^2=\left(\frac{\theta''}{\theta'}\right)^2\,.
\end{equation}
Taking the limit in which the degree of $\theta$ is large (i.e. for which $\theta:\text{S}^1\to\text{S}^1$ wraps the circle many times), we can approximate the above kinetic term via
\begin{equation}
\left(\frac{\theta''}{\theta'}\right)^2\sim-2\{\theta,\tau\}\,,
\end{equation}
where $\{\theta,\tau\}:=S[\theta](\tau)$ is the Schwarzian derivative of $\theta$ with respect to $\tau$. Therefore, the Nambu-Goto action of \eqref{eq:boundary-nambu-goto} can be written schematically as
\begin{equation}\label{eq:effective-schwarzian}
\int_{\partial\Sigma}\mathrm{d}\tau\,\{\theta,\tau\}\,,
\end{equation}
i.e. the dynamics are governed by a one-dimensional Schwarzian theory. Holographically, since we consider the covering space $\Sigma$ to be the worldsheet, the above action describes the motion of the endpoints of the string along the brane. The motion of the endpoints defines a function $\theta:\partial\Sigma\to\text{S}^1$, and the above action describes the effective dynamics of the string endpoints in the large-twist limit.\footnote{The Schwarzian action as the semiclassical Nambu-Goto action of an open string in $\text{AdS}_3$ was also recovered in \cite{Banerjee:2018twd,Banerjee:2018kwy}.}

Amazingly, the above action is precisely of the form of the action which defines the boundary dynamics of Jackiw-Teitelboim gravity in two dimensions \cite{Iliesiu:2019xuh}, thus hinting that, in this compactification limit, gravitational dynamics in the effective two-dimensional Euclidean theory has a subsector which is governed by a theory which formally resembles JT gravity. If we restore constants and take the central charge $c$ of the dual CFT to be generic, the effective action of the string endpoint is
\begin{equation}\label{eq:effective-schwarzian-constants}
S_{\text{eff}}=-\frac{c}{12\pi}\int\mathrm{d}\tau\,\{\theta,\tau\}\,.
\end{equation}
In JT gravity, the constant prefactor out front is given by $a/16\pi G_N^{(2)}$, where $G_N^{(2)}$ is the two-dimensional Newton constant and $a$ is a dimensionless parameter controlling the asymptotics of the diliton (see, for instance, Section 2.3.1 of \cite{Mertens:2022irh}). Relating this to the prefactor $c/12\pi$ above gives
\begin{equation}
G_N^{(2)}=\frac{3a}{4c}\,.
\end{equation}
Now, the 2D Newton constant in string theory is schematically related to the 3D Newton constant via $G_N^{(2)}= 2G_N^{(3)}\ell_s$, where $\ell_s$ is the string length. In the tensionless limit of string theory on $\text{AdS}_3$, the string length is equal to the $\text{AdS}_3$ radius $R$, and so
\begin{equation}
G_N^{(3)}=a\frac{3R}{2c}\,,
\end{equation}
which, up to the overall dimensionless constant $a$, is the famous Brown-Henneaux formula for the $\text{AdS}_3$ dual of a CFT with central charge $c$ \cite{Brown:1986nw}. Thus, not only does the effective action \eqref{eq:effective-schwarzian-constants} schematically reproduce the Schwarzian action for JT gravity, the constant prefactor can also be produced, assuming that the Newton constant $G_N^{(3)}$ is related to the central charge of the boundary theory by the Brown-Henneaux formula.\footnote{It is not clear however what the role of the genus-counting parameter $S_0$ of JT gravity is in this setup.}

There is a clear difference between \eqref{eq:effective-schwarzian} and the usual Schwarzian theory. In the context of JT gravity, the Schwarzian field $\theta:\text{S}^1\to\text{S}^1$ is a one-to-one diffeomorphism from the boundary circle onto itself, whereas in our case $\theta$ defines an $N$-to-one map. Such Schwarzian theories are considered in the literature, and are known to describe the dynamics of JT gravity with an $\mathbb{Z}_N$ conical defect inserted in the bulk \cite{Mefford:2020vde}. Retrospectively, the appearance of a conical defect in the effective 2D theory is not surprising, since it was argued in \cite{Eberhardt:2021jvj} and in Section \ref{sec:semiclassics} that, in the large-twist limit, the string generates a conical defect around any contractible cycle wound by the worldsheet.

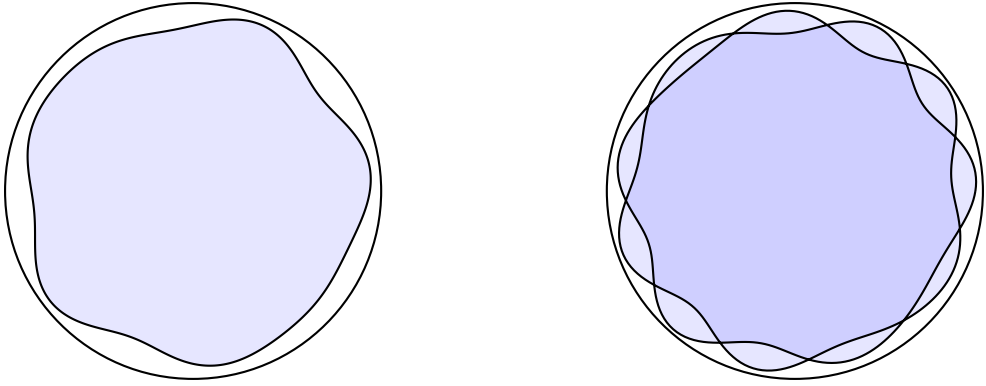
\begin{figure}
\centering
\begin{tikzpicture}
\begin{scope}[xshift = -4cm]
\draw[thick] (0,0) circle (2.5);
\fill[domain=0:360,scale=1,samples=500, blue, opacity = 0.1] plot ({\x}:{0.1*cos(5*\x) + 0.05*sin(7*\x) + 2.25});
\draw[thick, domain=0:360,scale=1,samples=500] plot ({\x}:{0.1*cos(5*\x) + 0.05*sin(7*\x) + 2.25});
\end{scope}
\begin{scope}[xshift = 4cm]
\draw[thick] (0,0) circle (2.5);
\fill[domain=0:360,scale=1,samples=500, blue, opacity = 0.1] plot ({\x}:{0.15*cos(5.5*\x) + 0.05*sin(7.5*\x) + 2.25});
\fill[domain=360:720,scale=1,samples=500, blue, opacity = 0.1] plot ({\x}:{0.15*cos(5.5*\x) + 0.05*sin(7.5*\x) + 2.25});
\draw[thick, domain=0:720,scale=1,samples=500] plot ({\x}:{0.15*cos(5.5*\x) + 0.05*sin(7.5*\x) + 2.25});
\end{scope}
\end{tikzpicture}
\caption{Left, a boundary cutoff in JT induced by the boundary diffeomorphism $f:\text{S}^1\to\text{S}^1$. Right, the boundary cutoff defined by the endpoint motion of a covering map $\Gamma:\partial\Sigma\to\text{S}^1$ with $\text{deg}(\Gamma)=2$.}
\label{fig:JT-winding}
\end{figure}

Actually, the existence of a relationship between a dimensional reduction of the tensionless string and the Schwarzian theory of JT gravity can be justified in another way. Let us consider JT gravity on the disk. The fundamental degrees of freedom of the theory are boundary gravitons, which correspond to diffeomorphisms of the boundary circle, up to a global M\"obius transformation. Such a large diffeomorphism can be expressed in terms of a function $\theta:\text{S}^1\to\text{S}^1$. In particular, one considers the partition function of JT gravity on the disk with a finite cut-off defined by a curve in the disk given by \cite{Engelsoy:2016xyb}
\begin{equation}\label{eq:JT-boundary-cutoff}
\theta=\theta(\tau)\,,\quad r=\varepsilon\frac{\mathrm{d}\theta}{\mathrm{d}\tau}(\tau)\,.
\end{equation}
Here, $r$ is the hyperbolic radial coordinate (i.e. $r\to\infty$ is the boundary of the disk). The parameter $\varepsilon$ is some infrared cutoff, which can be made arbitrarily small, so that the cutoff approaches the boundary. The Schwarzian theory is then considered by evaluating the JT action on the appropriate cutoff geometry, and then considering the path integral over $\text{Diff}(\text{S}^1)/\text{SL}(2,\mathbb{R})$.

The relationship to the string worldsheet in $\text{AdS}_3$ is that \eqref{eq:JT-boundary-cutoff} can be thought of as the projection of the classical worldsheet solution \eqref{eq:Wakimoto-classical} into Euclidean $\text{AdS}_2$. That is, we can consider $\theta$ to be the value of the covering map $\Gamma$ along the Euclidean $\text{AdS}_2$ brane in Figure \ref{fig:brane}. The primary difference, however, is that the worldsheet covering map is not a one-to-one map, but rather an $N$-to-one map (see Figure \ref{fig:JT-winding}).

The realisation of JT gravity as coming from the dimensional reduction of a brane setup in $\text{AdS}_3$ was also recently pointed out in \cite{Geng:2022slq,Geng:2022tfc,Deng:2022yll}. It is thus interesting to see that such a realisation seems to naturally arise in the context of the tensionless string theory on $\text{AdS}_3$. It would be worth exploring this relationship further to see if the Schwarzian form of the action \eqref{eq:effective-schwarzian} is truly an indication of some JT-like behaviour of strings in the presence of the spherical $\text{AdS}_2$ brane, or if it is simply a coincidence (perhaps due to the universality of the Schwarzian theory \cite{Ghosh:2019rcj}). This would likely require a much better understanding of the effective bulk field theory describing the tensionless string, which is still poorly understood.

\section{Summary and discussion}\label{sec:discussion}

\subsection{Summary}

In this paper, we explored the tensionless $\text{AdS}_3/\text{CFT}_2$ correspondence in the limit that all operator insertions have large twist with respect to the $\text{CFT}_2$ symmetric orbifold. We determined that, in this limit, one can recover the bulk geometry of the classical $\text{AdS}_3$ spacetime, despite the fact that naively the strings are `pinned' to the boundary sphere, and confirmed the postulate of \cite{Gaberdiel:2020ycd} that the Strebel differential can be recovered from the pullback of the $\text{AdS}_3$ metric to the worldsheet. We also discussed the relationship between the large-twist limit and the description of the $\text{AdS}_3$ worldsheet theory in terms of free twistorial fields, and in particular showed that the classical values of these fields obey a holomorphic Schr\"odinger equation whose potential is the Strebel differential. Finally, we considered the consequence of this analysis to the dynamics of a string which ends on a temporally localised brane, and found that the resulting worldsheet theory approximately reproduces the Schwarzian theory of JT gravity with conical singularities.

\subsection{Discussion}

The end goal of the analysis of this paper is to elucidate the bulk nature of tensionless string theory, which naively seems to have no bulk degrees of freedom. Because tensionless string theory is under such good analytic control, then, one would hope that one would be able to utilise this regime of the AdS/CFT correspondence to quantitatively and precisely explore questions in holography which are typically only accessible in the supergravity regime. We list a few of these potential directions below.

\subsection*{Black hole physics}

On of the great triumphs of holography is its ability to explain the microstate structure of (extremal) black holes through entropy calculations in the dual (strongly coupled) field theory \cite{Strominger:1996sh}. It would therefore be interesting to study Lorentzian black-hole geometries in $\text{AdS}_3$ in the tensionless limit, and determine if the typical entropic properties remain. Such black holes would necessarily be string-scale since, in the tensionless limit, the string length is of order of the spacetime geometry. It might also be possible to explore black holes as the extreme large-twist limit,\footnote{That is, the limit where the degree of the covering map $\Gamma$ is the same as the degree $K$ of the symmetric orbifold, which are then both taken to infinity.} in which the conical defects created by highly-twisted strings form a BTZ geometry \cite{Berenstein:2022ico}.

\subsection*{Information entropy and bulk reconstruction}

Recent developments in holography have attempted to reconstruct the bulk geometry of Anti-de Sitter spacetimes using information entropy in the dual CFT. In particular, the Ryu-Takayanagi (RT) formula relates the information entropy of a spacial region $\mathcal{A}$ to the minimal area of a codimension 2 surface $M(\mathcal{A})$ which extends into the $\text{AdS}$ bulk and which asymptotically ends on the boundary of $\mathcal{A}$ \cite{Ryu:2006ef}. The formula itself, however, is dependent on the existence of bulk semiclassical geometry, which in the tensionless string doesn't seem to exist aside from in the large-twist limit. Thus, it would be interesting to see if one could reproduce something analogous to the RT entropy in this limit.

\vspace{0.5cm}

Apart from the application of the large twist limit to more `classical' questions of holography, there are generalisations of this construction which would be interesting to explore.

\subsection*{Extension to higher dimensions} 

Just as in $\text{AdS}_3\times\text{S}^3\times\mathcal{M}$, there also exists an analogous tensionless description of string theory on $\text{AdS}_5\times\text{S}^5$ whose worldsheet spectrum precisely reproduces the single-trace spectrum of $\mathcal{N}=4$ super Yang-Mills in four dimensions \cite{Gaberdiel:2021qbb,Gaberdiel:2021jrv}. Moreover, an analogous construction for correlation functions to that of the symmetric orbifold in two dimensions seems to exist \cite{Bhat:2021dez}, with an honest worldsheet derivation currently under exploration \cite{Gaberdiel:2022jkl}. Since the analogue of the twist $w$ in SYM is the number of letters in a single trace operator $\text{Tr}(\Phi_1\cdots\Phi_w)$, one could hope, then, that a similar geometric picture in terms of some generalisation of Strebel differentials could arise in $\mathcal{N}=4$ SYM in the limit that the operators involved are traces of very many letters. Such an analysis would be very similar to the so-called `large $p$' analysis of \cite{Aprile:2020luw} to the limit of vanishing 't Hooft coupling.

\subsection*{Deforming away from the orbifold point}

The tensionless limit of string theory is an unusually well-behaved holographic theory because its dual CFT is free. In order to understand less well-behaved examples of AdS/CFT, it would be useful to move beyond the tensionless limit, i.e. worldsheet theories with $k>1$ units of NS-NS flux. The proposed CFT dual is given by a symmetric orbifold CFT deformed by a certain twist-2 field $\sigma_2$ \cite{Eberhardt:2021vsx}. In conformal perturbation theory, the perturbation of correlation functions $\Braket{\prod_{i}V^{w_i}}$ is computed schematically by considering correlators $\int\Braket{(\sigma_2)^k\prod_{i}V^{w_i}}$, where the positions of the $\sigma_2$ insertions are integrated over.\footnote{More precisely, the deforming operator is a $G_{-1/2}$ descendant of the twice spectrally-flowed vacuum $V^{(2)}$.} Such a correlator will be represented by a covering map $\Gamma_k$ which has usual critical points of order $w_i$ at $z_i$, as well as $k$ `extra' simple critical points at the insertions of the deformation operators. In general, constructing a covering map $\Gamma_k$ from the original covering map $\Gamma_0$ (without deformation operators) is a very tricky task. However, in the large-twist limit, we can think of adding a simple critical point to the covering map as a kind of perturbation in $1/\text{deg}(\Gamma)$. In particular, it would be interesting to explore how the Strebel differential $\varphi_k$ associated to the covering map $\Gamma_k$ is related to the un-deformed differential $\varphi_0$.\footnote{The number of such coverings is given by the ELSV formula \cite{Ekedahl:1999}, which is written in terms of integrals of certain characteristic classes over the compactified moduli space $\overline{\mathcal{M}}_{g,n}$, and thus an understanding of the ELSV formula at large-twist would potentially be necessary.} This would potentially allow for an analytic check of the duality for $k>1$ strings \cite{Eberhardt:2021vsx} in the large-twist limit.

\subsection*{Ensemble averaging}

Another way to explore the bulk geometry in more detail would be to consider an ensemble average of symmetric orbifold theories. In particular, one can average quantities in the symmetric orbifold theory $\text{Sym}^K(\mathbb{T}^4)$ over the Narain moduli space of 2D sigma models with target space $\mathbb{T}^4$.\footnote{Averaging the symmetric orbifold theory was already discussed in \cite{Eberhardt:2021jvj}, however the resulting bulk theory has not yet been explored in detail.} In the non-orbifolded case, the resulting averaged partition function has an interpretation in terms of a $\text{U}(1)^4\times\text{U}(1)^4$ Chern-Simons theory coupled to an exotic topological theory of gravity which performs a sum over bulk geometries \cite{Maloney:2020nni}. Moreover, Narain-averaged orbifolds of the form $\mathbb{T}^D/\mathbb{Z}_N$ have been considered, and their interpretation includes an orbifolded Chern-Simons theory coupled to an even more exotic theory of gravity which sums over bulk geometries with codimension-2 orbifold defects \cite{Benjamin:2021wzr}, whose topologies are restricted to so-called `rational tangles' (see also \cite{Maloney:2016kee}). In the symmetric orbifold case, one could expect a bulk interpretation in terms of a supersymmetric Chern-Simons theory whose gauge group contains the discrete orbifold group $S_K$ as a factor. Indeed, such a bulk interpretation is currently under consideration \cite{Kames-King:2022xyz}.

\subsection*{Relation to JT gravity}

In Section \ref{sec:ads2} we hinted at a potential relationship between a compactification of the tensionless string to euclidean $\text{AdS}_2$ and JT gravity in two dimensions. In particular, we found that the effective action of the endpoints of the string are described by the Schwarzian derivative of the endpoint coordinates, which is identical to the effective action of boundary gravitons in JT gravity. It would be interesting to study this relationship further. Since JT gravity is known to be dual to a matrix model \cite{Saad:2019lba}, it would be particularly interesting to study the large-twist limit of the symmetric orbifold CFT with boundary and see if there exists of a subsector of the symmetric orbifold which recovers the matrix model of JT gravity.

\subsection*{Acknowledgements}

I would like to thank Matthias Gaberdiel, Rajesh Gopakumar, Andrea Dei, Francesco Galvagno, and Pronobesh Maity for helpful conversations and comments on an early version of the manuscript. I would also like to thank Thomas Merthens, Joshua Kames-King, Mykhaylo Usatyuk, Jakub Vo\v{s}mera, Shota Komatsu, and Lorenz Eberhardt for useful discussions. Finally, I would like to thank the organisers of the SwissMAP workshop ``Deciphering the AdS/CFT correspondence,'' during which many of these ideas were developed and refined. This work was supported by the Swiss National Science Foundation through a personal grant and via the NCCR SwissMAP.

\appendix

\section{The tensionless worldsheet theory}\label{sec:free-fields}

In this appendix, we review the free field construction of \cite{Eberhardt:2018,Dei:2020} for the tensionless string theory on $\text{AdS}_3\times\text{S}^3\times\mathbb{T}^4$, focusing almost exclusively on the bosonic sector which generates $\text{AdS}_3$. We also demonstrate that an equivalent worldsheet theory, related to the `old' theory by a twist in the stress-tensor, leads to a more natural definition of correlation functions, and allows for the realisation of the worldsheet free fields as solutions to the holomorphic Schr\"odinger equation considered in Section \ref{sec:reconstruction}.

\subsection[The free field construction of \texorpdfstring{$\mathfrak{psu}(1,1|2)_1$}{psu(1,1|2)1}]{\boldmath The free field construction of \texorpdfstring{$\mathfrak{psu}(1,1|2)_1$}{psu(1,1|2)1}}

We consider the free field realisation consisting of two pairs $(\lambda,\mu^{\dagger})$ and $(\mu,\lambda^{\dagger})$ of $h=\frac{1}{2}$ free fields on the worldsheet satisfying the OPEs\footnote{Relative to the conventions of \cite{Dei:2020,Knighton:2020kuh}, we have defined $\lambda=\xi^+$, $\mu=-\xi^-$, $\lambda^{\dagger}=\eta^+$, and $\mu^{\dagger}=\eta^-$. This is to make our notation closer to that of \cite{Gaberdiel:2021qbb}.}
\begin{equation}
\lambda(z)\mu^{\dagger}(w)\sim\frac{1}{z-w}\,,\quad\mu(z)\lambda^{\dagger}(w)\sim\frac{1}{z-w}\,.
\end{equation}
This defines two $\lambda=\frac{1}{2}$ $\beta\gamma$ systems, for which we can write the action
\begin{equation}
S=\int(\lambda\bar\partial\mu^{\dagger}+\mu\bar\partial\lambda^{\dagger})
\end{equation}
as well as the stress tensor
\begin{equation}
T=\frac{1}{2}\left(\mu^{\dagger}\partial\lambda-\lambda\partial\mu^{\dagger}+\lambda^{\dagger}\partial\mu-\mu\partial\lambda^{\dagger}\right)\,.
\end{equation}
Bilinears in the fields $(\lambda,\mu,\lambda^{\dagger},\mu^{\dagger})$ generate the Kac-Moody algebra $\mathfrak{sp}(4)_1$ \cite{Gaiotto:2017euk}. The currents
\begin{equation}
J^+=\lambda^{\dagger}\lambda\,,\quad J^-=\mu^{\dagger}\mu\,,\quad J^3=\frac{1}{2}\left(\lambda^{\dagger}\mu-\mu^{\dagger}\lambda\right)
\end{equation}
generate the subalgebra $\mathfrak{sl}(2,\mathbb{R})_1\subset\text{sp}(4)_1$, while the current
\begin{equation}
U=\frac{1}{2}\left(\lambda^{\dagger}\mu+\mu^{\dagger}\lambda\right)
\end{equation}
generates a $\mathfrak{u}(1)$ orthogonal to the $\mathfrak{sl}(2,\mathbb{R})$. Gauging the free field theory by the current $U$ leaves only the $\mathfrak{sl}(2,\mathbb{R})$ degrees of freedom.

If we think of the free fields as two $\beta\gamma$ systems, then each comes equipped with a conserved current corresponding to the $U(1)$ symmetry of the OPEs. These currents are given by
\begin{equation}
J_1=\lambda^{\dagger}\mu\,,\quad J_2=\mu^{\dagger}\lambda\,.
\end{equation}
Thus, the currents $J^3$ and $U$ are given by
\begin{equation}
J^3=\frac{1}{2}(J_1-J_2)\,,\quad U=\frac{1}{2}(J_1+J_2)\,.
\end{equation}
Due to the fact that the $\beta\gamma$ systems have conformal weight $\lambda=\frac{1}{2}$, neither of the currents $J_1,J_2$ are anomalous.

\subsection{Twisting the free field realisation}

The realisation above certainly has advantages, specifically the non-anomalous behaviour of the currents $J^3,U$. However, as we have seen through the investigation of correlation functions, it also has its problems. In particular, it seems to be the case \cite{Dei:2020} that in order to define non-vanishing correlators, one has to introduce $n+2g-2$ auxiliary fields $W$, whose physical significance is unclear.

It turns out that there is an alternative formulation of the $\mathfrak{sl}(2,\mathbb{R})_1$ theory which eliminates these problems and which, from the point of view of the correlation functions, leads to much more natural results. We accomplish this by `twisting' the theory via the new stress tensor\footnote{In the full supersymmetric theory, we instead twist by adding the term $\partial Z$, where $Z$ is the current defined in \cite{Dei:2020}. Since $Z$ is itself a null-current, this twisting will have no effect on the central charge of the system.}
\begin{equation}
T'=T+2\partial U\,.
\end{equation}
Since physical states in the theory will satisfy $U_n\ket{\varphi}=0$ for all $n\geq 0$, this twisting can not have an effect on the spectrum. However, it does have the effect of changing the conformal weights of the free fields. In particular, the new conformal weights are
\begin{equation}
\begin{split}
\lambda,\mu\,,\quad h=-\frac{1}{2}\,,\\
\lambda^{\dagger},\mu^{\dagger}\,,\quad h=\frac{3}{2}\,.
\end{split}
\end{equation}
Thus, the twisting effectively results in considering two $\beta\gamma$ systems with $\lambda=-\frac{1}{2}$ as opposed to $\lambda=\frac{1}{2}$. Indeed, the new stress tensor is
\begin{equation}
T'=\frac{1}{2}\left(\mu^{\dagger}\partial\lambda-\lambda\partial\mu^{\dagger}+\lambda^{\dagger}\partial\mu-\mu\partial\lambda^{\dagger}\right)+\partial(\lambda^{\dagger}\mu)+\partial(\mu^{\dagger}\lambda)\,,
\end{equation}
which is precisely the stress tensor of two $\lambda=-\frac{1}{2}$ $\beta\gamma$ theories.

The effect of this twist is that not all of the currents are non-anomalous. For example, the OPE between the stress-tensor and the current $U$ is given by
\begin{equation}
\begin{split}
T'(z)U(w)&\sim\frac{1}{(z-w)^2}+\frac{\partial U(w)}{z-w}+2\partial U(z)U(w)\\
&\sim-\frac{2}{(z-w)^3}+\frac{1}{(z-w)^2}+\frac{\partial U(w)}{z-w}\,,
\end{split}
\end{equation}
and so the current $U$ is no longer a primary field of weight $h=1$, but instead carries a `background charge' of $Q=-2$. Practically, this implies that any correlation function
\begin{equation}
\Braket{\prod_{i}\varphi_i}
\end{equation}
vanishes unless the total $U$-charge carried by the states $\varphi_i$ is $2-2g$. As we will see below, this actually becomes an advantage when computing correlation functions.

\subsection{Highest-weight representations}

Defining states in the twisted free field theory is almost identical to the untwisted theory, up to a couple of subtleties. Recall that for a field $\varphi$ of conformal weight $h$, the NS-vacuum $\ket{\Omega}$ satisfies
\begin{equation}
\varphi_r\ket{\Omega}=0\,,\quad \forall r\geq 1-h\,.
\end{equation}
Thus, for our free fields, we have
\begin{equation}
\begin{split}
\lambda_r\ket{\Omega}=\mu_r\ket{\Omega}=0\,,&\quad\forall r\geq\frac{3}{2}\,,\\
\lambda^{\dagger}_r\ket{\Omega}=\mu^{\dagger}_r\ket{\Omega}=0\,,&\quad\forall r\geq-\frac{1}{2}\,.
\end{split}
\end{equation}
Furthermore, the Ramond sector similarly satisfies
\begin{equation}
\begin{split}
\lambda_r\ket{\varphi}=\mu_r\ket{\varphi}=0\,,&\quad\forall r\geq 2\,,\\
\lambda^{\dagger}_r\ket{\varphi}=\mu^{\dagger}_r\ket{\varphi}=0\,,&\quad\forall r\geq 0\,.
\end{split}
\end{equation}
The `zero modes' of the twisted free fields are identified with $\lambda_1,\mu_1$ and $\lambda^{\dagger}_{-1},\mu^{\dagger}_{-1}$. We can define the $J^3$ quantum number $m$ and the $U$ quantum number $j-\frac{1}{2}$, which generate the Cartan of the free field realisation. With respect to this basis, the bottom of the Ramond representation takes the form
\begin{equation}
\begin{split}
\lambda_1\ket{m,j}=\ket{m+\tfrac{1}{2},j-\tfrac{1}{2}}\,,&\quad\mu^{\dagger}_{-1}\ket{m,j}=-(m-j)\ket{m-\tfrac{1}{2},j+\tfrac{1}{2}}\,,\\
\mu_1\ket{m,j}=\ket{m-\tfrac{1}{2},j-\tfrac{1}{2}}\,,&\quad\lambda^{\dagger}_{-1}\ket{m,j}=(m+j)\ket{m+\tfrac{1}{2},j+\tfrac{1}{2}}\,.
\end{split}
\end{equation}
Defining the vertex operators $V_{m,j}(z)$ in the obvious way, we can read off the OPE
\begin{equation}
\lambda(z)V_{m,j}(0)\sim\mathcal{O}(z^{-\frac{1}{2}})
\end{equation}
and similarly for all other fields.

\subsection{Spectral flow and picture number}

Spectral flow is defined precisely as before, in terms of two operators $\sigma^{(+)}$ and $\sigma^{(-)}$ which separately spectrally flow the two $\beta\gamma$ systems, namely
\begin{equation}
\sigma^{(+)}(\mu_r)=\mu_{r+\frac{1}{2}}\,,\quad\sigma^{(+)}(\lambda^{\dagger}_r)=\lambda^{\dagger}_{r-\frac{1}{2}}
\end{equation}
and
\begin{equation}
\sigma^{(-)}(\lambda_r)=\lambda_{r-\frac{1}{2}}\,,\quad\sigma^{(-)}(\mu^{\dagger}_r)=\mu^{\dagger}_{r+\frac{1}{2}}\,.
\end{equation}
It should also be noted that if we bosonise the $U(1)$ currents $J_1=\partial\phi_1$ and $J_2=\partial\phi_2$, then the spectral flow operators are simply
\begin{equation}
\sigma^{(+)}=e^{\phi_1/4}\,,\quad\sigma^{(-)}=e^{-\phi_2/4}\,.
\end{equation}
Furthermore, we can consider the combinations $\sigma=\sigma^{(+)}\sigma^{(-)}$ and $\widehat{\sigma}=\sigma^{(+)}(\sigma^{(-)})^{-1}$, or equivalently, if we bosonise $J^3=\partial\phi$ and $U=\partial\beta$, we have
\begin{equation}
\sigma=e^{\phi/2}\,,\quad\widehat{\sigma}=e^{\beta/2}\,.
\end{equation}
Indeed, $\sigma$ raises the $J^3$ eigenvalue by $1/2$ and $\widehat{\sigma}$ raises the $U$ eigenvalue by $1/2$.

Given a state $\varphi$ in the Ramond sector, we can define an array of spectrally flowed states by acting with $\sigma$ and $\widehat{\sigma}$. Specifically, we define
\begin{equation}
\varphi^{(p,q)}=\sigma^p\circ\widehat{\sigma}^q\left(\varphi\right)=e^{\phi p/2}e^{\beta q/2}\varphi=e^{\beta q/2}\sigma^w(\varphi)\,.
\end{equation}
Due to the analogy with the usual $\beta\gamma$ superdiffeomorphism ghost system, we will refer to the spectral flow in the $\widehat{\sigma}$ direction as the \textit{picture number} of the state (a similar observation was made in \cite{Gaberdiel:2021njm}, c.f. equation (2.4)).\footnote{Note that the current $U$ is defined with a relative minus sign to the standard $\beta\gamma$ current, and so the picture number also carries a relative minus sign, as well as a relative factor of two since $\widehat{\sigma}$ only carries a $U$-charge of $1/2$.}

Given a HW state in the $(p,q)$ sector, we have the mode relations
\begin{equation}
\begin{split}
\lambda_r\ket{\varphi}^{(p,q)}&=0\,,\quad\forall r\geq 2+\frac{p-q}{2}\\
\mu_r\ket{\varphi}^{(p,q)}&=0\,,\quad\forall r\geq 2-\frac{p+q}{2}\\
\lambda^{\dagger}_r\ket{\varphi}^{(p,q)}&=0\,,\quad\forall r\geq\frac{p+q}{2}\\
\mu^{\dagger}_r\ket{\varphi}^{(p,q)}&=0\,,\quad\forall r\geq-\frac{p-q}{2}\,.
\end{split}
\end{equation}
It should be noted that the $q=2$ picture essentially recovers the original formulation of the Ramond sector. Indeed, consider $p=0$ and $q=2$. Then the above mode relations all coincide to say that $R$-sector ground states in the $q=2$ picture are annihilated by only the positive modes of the free fields, and the zero modes move the states around in the representation. As we will see later, the $q=2$ picture number is also the `natural' one to consider for defining correlation functions.

For completeness, we note the OPEs
\begin{equation}
\begin{split}
\lambda(z)\varphi^{(p,q)}(0)\sim z^{\frac{q-p-1}{2}}\,,&\quad\mu(z)\varphi^{(p,q)}(0)\sim z^{\frac{q+p-1}{2}}\,,\\
\lambda^{\dagger}(z)\varphi^{(p,q)}(0)\sim z^{-\frac{p+q+1}{2}}\,,&\quad\mu(z)\varphi^{(p,q)}(0)\sim z^{\frac{p-q-1}{2}}\,.\\
\end{split}
\end{equation}
Specifically, for $p=w$ and $q=2$, we have
\begin{equation}\label{eq:OPEs}
\begin{split}
\lambda(z)\varphi^{(w,2)}(0)\sim z^{-\frac{w-1}{2}}\,,&\quad\mu(z)\varphi^{(w,2)}(0)\sim z^{\frac{w+1}{2}}\,,\\
\lambda^{\dagger}(z)\varphi^{(w,2)}(0)\sim z^{-\frac{w+3}{2}}\,,&\quad\mu^{\dagger}(z)\varphi^{(w,2)}(0)\sim z^{\frac{w-3}{2}}\,.
\end{split}
\end{equation}

\subsection{Correlation functions}

Now that we have discussed spectral flow and representations of the free field theory, we are in position to define correlation functions. The following treatment will be somewhat imprecise. In particular, the full definition of correlation functions requires the technology of the hybrid formalism of Berkovits, Vafa, and Witten \cite{Berkovits_1999}, see also Section 3 of \cite{Dei:2020}. Here we will only review the salient features.

In the unflowed Ramond sector, physical states satisfy the condition $U_0\ket{m,j}=0$, or $j=\frac{1}{2}$. However, as we know, in order to define a correlation function in the hybrid formalism, we need to include $n+2g-2$ PCOs, whose overall effect (from the Bosonic viewpoint) is lowering the value of $j$ by one. Thus, let $\tilde{j}$ be the value of $j$ \textit{after} picture changing. Then the correlator
\begin{equation}
\Braket{\prod_{i=1}^{n}V_{m_i,\tilde{j}_i}^{w_i}(x_i,z_i)}
\end{equation}
satisfies
\begin{equation}
\sum_{i=1}^{n}\left(\tilde{j}_i-\frac{1}{2}\right)=-(n+2g-2)\,.
\end{equation}
However, such a correlator must vanish, since the sum of all $U$-charges must be $2-2g$. We can remidy this, however, by artificially increasing the $U$-charge of all $n$ states by $1$. The simplest way to do this is to consider states in the $q=2$ picture. That is, we define
\begin{equation}
\widetilde{V}^{w}_{m,j}=V^{(w,2)}_{m,j}=e^{\beta}V^{w}_{m,j}\,.
\end{equation}
Such a state is indistinguishable from $V_{m,j}^{w}$ at the level of the $\mathfrak{sl}(2,\mathbb{R})$ theory, but now carries a $U$-charge $j+\frac{1}{2}=(j-\frac{1}{2})+1$. Thus, the correlator
\begin{equation}
\Braket{\prod_{i=1}^{n}\widetilde{V}_{m_i,\tilde{j}_i}^{w_i}(x_i,z_i)}
\end{equation}
satisfies
\begin{equation}
\sum_{i=1}^{n}U_i=\sum_{i=1}^{n}\left(\tilde{j}+\frac{1}{2}\right)=2-2g\,,
\end{equation}
and thus is generically nonzero. Furthermore, note that this charge conservation condition can be written as
\begin{equation}
\sum_{i=1}^{n}\tilde{j}_i=2-2g-\frac{n}{2}=\frac{1}{2}(n+2g-2)-(n+3g-3)\,,
\end{equation}
which is precisely the so-called `$j$-constraint' of \cite{Eberhardt:2019} required for $\mathfrak{sl}(2,\mathbb{R})_1$ correlation functions to have a localising solution. Note, furthermore, that we were able to define such a correlation function without introducing the $W$-fields of \cite{Dei:2020}.\footnote{Strictly speaking, our construction is identical to placing a $W$-field at the location $z=z_i$ for each vertex operator. The difference between this approach and that of \cite{Dei:2020} is that there are only $n$ such insertions and that there are canonical locations, i.e. the locations of these operators are not free parameters.}

\subsection{Localisation}

We now argue that the worldsheet correlators in the free field theory localize to covering maps from the worldsheet to $\mathbb{CP}^1$. For now, let us keep $j_i$ generic, taking them to their critical values only as a last step, and consider the sections of $S^{-1}$ (the bundle of half-differentials $f(z)(\mathrm{d}z)^{-1/2}$ on the worldsheet)

\begin{equation}
L(z)=\Braket{\lambda(z)\prod_{i=1}^{n}\widetilde{V}^{w_i}(x_i,z_i)}\,,\quad M(z)=\Braket{\mu(z)\prod_{i=1}^{n}\widetilde{V}^{w_i}(x_i,z_i)}\,.
\end{equation}
By the OPEs in \eqref{eq:OPEs}, we see that $L$ and $M$ both have poles of order $\frac{w_i-1}{2}$ at $z=z_i$. Thus, we have
\begin{equation}\label{eq:divisors}
\text{div}(L)=-\sum_{i=1}^{n}\frac{w_i-1}{2}z_i+\Delta^-\,,\quad\text{div}(M)=-\sum_{i=1}^{n}\frac{w_i-1}{2}z_i+\Delta^+\,,
\end{equation}
where $\Delta^{\pm}$ are divisors of order $1-g+\sum_{i=1}^{n}\frac{w_i-1}{2}=N$. Thus, $M/L$ is a function of order $N$ which satisfies
\begin{equation}
\frac{M}{L}(z)-x_i\sim(z-z_i)^{w_i}
\end{equation}
at all $z=z_i$. Thus, $M/L=\Gamma$ is the covering map. Furthermore, we see that $\Delta^-$ is the divisor of poles of $\Gamma$ and $\Delta^+$ is the divisor of zeroes. The only sections of $S^{-1}$ with divisors \eqref{eq:divisors} are
\begin{equation}
L\propto\frac{1}{\sqrt{\partial\Gamma}}\,,\quad M\propto\frac{\Gamma}{\sqrt{\partial\Gamma}}\,,
\end{equation}
up to an overall constant.

\subsection{The daggered fields}

We have now shown that the correlators of this theory localise by using the analyticity properties of the undaggered half of the symplectic Boson theory. It is also instructive to look at the daggered fields. However, for these fields, analyticity and Ward identities are not enough to fully narrow down their behaviour. The primary reason for this is that these fields have conformal weight $h=\frac{3}{2}$, and thus their correlation functions are sections of a line bundle with higher degree than that of the $(\lambda,\mu)$ fields. If we consider the correlators given by
\begin{equation}
L^{\dagger}(z)=\Braket{\lambda^{\dagger}(z)\prod_{i=1}^{n}\widetilde{V}^{w_i}(x_i,z_i)}\,,\quad M^{\dagger}(z)=\Braket{\mu^{\dagger}(z)\prod_{i=1}^{n}\widetilde{V}^{w_i}(x_i,z_i)}\,,
\end{equation}
we see that $L^{\dagger},M^{\dagger}$ have poles of order $\frac{w_i+3}{2}$, and so the divisors
\begin{equation}
\Delta^-=\text{div}(L^{\dagger})+\sum_{i=1}^{n}\frac{w_i+3}{2}z_i\,,\quad\Delta^+=\text{div}(M^{\dagger})+\sum_{i=1}^{n}\frac{w_i+3}{2}z_i\,,
\end{equation}
have degree
\begin{equation}
\text{deg}(\Delta^{\pm})=\sum_{i=1}^{n}\frac{w_i+3}{2}+3g-3=N+2(n+2g-2)\,.
\end{equation}
That is, the fields $L^{\dagger},M^{\dagger}$ have $2(n+2g-2)$ extra degrees of freedom each, which cannot be constrained by the Ward identities.

Another way to see this is to consider the definitions
\begin{equation}
L^{\dagger}=\frac{\phi_1}{\sqrt{\partial\Gamma}}\,,\quad M^{\dagger}=\frac{\Gamma\phi_2}{\sqrt{\partial\Gamma}}\,,
\end{equation}
where $\phi_1,\phi_2\in H^0(K^2,\Sigma)$ are quadratic differentials on the worldsheet. In order for $L^{\dagger},M^{\dagger}$ to have poles of order $\frac{w_i+3}{2}$ at $z=z_i$, we need that $\phi_1,\phi_2$ have double poles at those locations. That is,
\begin{equation}
\phi_1,\phi_2\sim\frac{(\mathrm{d}z)^2}{(z-z_i)^2}+\cdots\,.
\end{equation}
In terms of divisors, we can define
\begin{equation}
D^-=\text{div}(\phi_1)+2\sum_{i=1}^{n}z_i\,,\quad D^+=\text{div}(\phi_2)+2\sum_{i=1}^{n}z_i\,,
\end{equation}
which have degree
\begin{equation}
\text{deg}(D^{\pm})=2(n+2g-2)\,.
\end{equation}
That is, the `extra' degrees of freedom we cannot solve for are contained in the quadratic differentials $\phi_1,\phi_2$.

\subsection{Including fermions}

Finally, for completion, we describe how the full supersymmetric theory works. We include, in addition to the bosonic fields $(\lambda,\mu)$ and $(\mu^{\dagger},\lambda^{\dagger})$ a pair of free fermions $\psi^a$ of weight $h=-\frac{1}{2}$ and their canonical conjugates $(\psi^{\dagger})_a$ of weight $h=\frac{3}{2}$ with $a=1,2$. Specifically, we have the OPEs
\begin{equation}
\psi^a(z)(\psi^{\dagger})_b(w)\sim\frac{\delta\indices{^a_b}}{z-w}\,.
\end{equation} 
The chiral half of the theory has the action
\begin{equation}
S=\int(\lambda\bar\partial\mu^{\dagger}+\mu\bar\partial\lambda^{\dagger}+\psi^a\bar{\partial}(\psi^{\dagger})_a)\,.
\end{equation}
The central charge of the full theory is given by $c=0$, since the bosonic and fermionic parts of the theory are identical except for their statistics. The $\mathfrak{psu}(1,1|2)_1$ model is then obtained by gauging the symmetry
\begin{equation}
\begin{split}
(\lambda,\mu)\to\alpha(\lambda,\mu)\,,&\quad (\mu^{\dagger},\lambda^{\dagger})\to\alpha^{-1}(\mu^{\dagger},\lambda^{\dagger})\,,\\
\psi^a\to\alpha\,\psi^a\,,&\quad(\psi^{\dagger})_a\to\alpha^{-1}(\psi^{\dagger})_a\,.
\end{split}
\end{equation}
This symmetry is generated by the current
\begin{equation}
Z=\frac{1}{2}\left(\lambda^{\dagger}\mu+\mu^{\dagger}\lambda+\psi^a(\psi^{\dagger})_a\right)\,,
\end{equation}
and the construction defined here is related to that of \cite{Eberhardt:2018,Dei:2020} via the twist
\begin{equation}
T_{\text{new}}=T_{\text{old}}+2\partial Z\,.
\end{equation}
Since $Z$ is null, this twist does not change the conformal weight of the gauged $\mathfrak{psu}(1,1|2)_1$ worldsheet theory, and thus it is still critical once one includes the $\mathbb{T}^4$ and ghost contributions. Unlike in the bosonic case, the conservation of $Z$ is not anomalous, but the conservation of the conjugate current
\begin{equation}
Y=\frac{1}{2}\left(\lambda^{\dagger}\mu+\mu^{\dagger}\lambda-\psi^a(\psi^{\dagger}_a)\right)
\end{equation}
is anomalous.

\bibliography{draft1}
\bibliographystyle{utphys.bst}

\end{document}